\documentclass[11pt,a4paper]{article}
\usepackage{jinstpub}
\usepackage{lineno}

\title{Modeling of Advanced Accelerator Concepts}
\author[a]{J.-L. Vay}
\author[a]{A. Huebl}
\author[a]{R. Lehe}
\author[b]{N.M. Cook}
\author[c] {R.~J. England}
\author[d] {U. Niedermayer}
\author[e,f]{P. Piot}
\author[g]{F. Tsung}
\author[h]{D. Winklehner}

\affiliation[a]{Lawrence Berkeley National Laboratory, Berkeley, CA 94720, USA}
\affiliation[b]{RadiaSoft LLC, RadiaSoft LLC, Boulder, CO 80301 USA}
\affiliation[c]{SLAC National Accelerator Laboratory, Menlo Park, CA, 94025, USA}
\affiliation[d] {Technical University Darmstadt, 64289 Darmstadt, Germany}
\affiliation[e]{Northern Illinois University, DeKalb, IL 60115 USA}
\affiliation[f]{Argonne National Laboratory, Lemont, IL 60439 USA}
\affiliation[g]{University of California Los Angeles, Los Angeles, CA 90095, USA}
\affiliation[h]{Massachusetts Institute of Technology, Cambridge, MA, 02139, USA}

\emailAdd{jlvay@lbl.gov}

\date{today}

\abstract{Computer modeling is essential to research on Advanced Accelerator Concepts (AAC), as well as to their design and operation. This paper summarizes the current status and future needs of AAC systems and reports on several key aspects of (i) high-performance computing (including performance, portability, scalability, advanced algorithms, scalable I/Os and In-Situ analysis), (ii) the benefits of ecosystems with integrated workflows based on standardized input and output and with integrated frameworks developed as a community, and (iii) sustainability and reliability (including code robustness and usability).}
\begin{document}

\maketitle

\section{Introduction}

Advanced Accelerator Concepts (AAC) can produce accelerating or focusing gradients that are orders of magnitude higher than the  gradients in conventional accelerator technologies. Thus they have the potential to enable much more compact - and in some cases cheaper - accelerator facilities. AAC encompass several technologies, including laser-driven plasma acceleration (a.k.a. LWFA or LPA), beam-driven plasma acceleration (a.k.a. PWFA), structure-based wakefield accelerators (SWFA), dielectric laser acceleration (DLA), laser-ion acceleration and novel plasma devices (e.g., capillary discharge plasmas and laser-ionized plasma columns). Efficiently modeling these different concepts require different types of simulation tools. DLA, for instance, can be mostly modeled with simulation tools that are already in use for conventional accelerators (e.g. CST~\cite{CST} or VSim~\cite{VSIM}). On the other hand, plasma-based schemes require different simulation tools, and can be considerably more computationally expensive. In this paper, we describe the specific needs and challenges that apply when modeling advanced accelerator concepts. We note that some of the challenges of modeling conventional accelerators, which are described in another paper of this issue \cite{SaganICFA2021}, also apply to advanced accelerator concepts. This includes for instance software sustainability, standardization, validation, verification, usability, integrated workflows and frameworks, toolkits, ecosystems, and future computing architectures.

Plasma acceleration involves a laser or charged particle beam ``driver'' that displaces electrons as it propagates through a narrow and long (in comparison to the dimensions of the beam) channel of pre-ionized plasma \cite{ChenPRL1985,Esareyrmp09}. The displaced electrons create a pocket of positively charged protons (which move very little thanks to their inertia) that attracts the electrons back on the axis of propagation of the driver, initiating plasma oscillations that form a ``wake'' of very intense, alternately positive and negative, electric fields that follow the driver beam. These fields can be used to accelerate and guide a ``witness'' electron or positron beam to very high energy in a much shorter distance than is possible using conventional accelerating techniques. 
Several aspects of plasma accelerators are particularly challenging to model.  These include the detailed kinetic modeling (going beyond the fluid approximation by following plasma particle trajectories) of trapping of background plasma electrons in the wakefields, ion motion,  the propagation of ultra-low emittance beams in the plasma, the long time (ps to ns) evolution of the perturbed plasma, and, because of the large number of 3D simulations required, evaluation of tolerances to non-ideal effects.
It is also essential to integrate physics models beyond single-stage plasma physics in the current numerical tools, incorporating: plasma hydrodynamics for long term plasma/gas evolution and accurate plasma/gas profiles, collision, ionisation \& recombination, coupling with conventional beamlines, production of secondary particles, spin polarization, QED physics at the interaction point.

Structure-based wakefield accelerators (SWFA) exploit the complex frequency dependence of dielectric and meta-material structures to selectively excite electromagnetic modes for the acceleration and control of energetic electron beams. As with plasma accelerators, the careful arrangement of a driver and witness beam can permit the acceleration of electron beams with GV/m-scale gradients~\cite{OShea_2020}. Current research efforts aim to improve the efficiency and stability of these interactions through the manipulation of both the driver and structure properties. For example, the longitudinal shaping of the driver can enhance the resulting transformer ratio, improving the maximum energy gain of the witness beam for a given driver energy~\cite{Gao_2018}. Alternatively, modifications to the structure composition and boundaries may improve the central frequency, bandwidth, and phase-velocity of the excited modes~\cite{Andonian_2014, Hoang_2018,Lemery_2018,Lu_2019}. The modeling of these complex boundary conditions imposes additional requirements on core algorithms to capture the electromagnetic response at these boundaries without introducing numerical errors, while also demanding high level tools to describe and initialize representations of a structure within the simulation domain.

The dielectric laser-driven accelerator (DLA) uses a short-pulse infrared (IR) laser, typically in the 0.8 to 2 micron wavelength range, to drive an electromagnetic accelerating mode inside the narrow vacuum channel of a periodic dielectric device \cite{england:rmp:2014}. The combination of infrared frequencies, femtosecond pulses, and dielectric materials possessing high damage thresholds enables GV/m accelerating fields \cite{cesar:nonlinear:2018}. Similar to conventional radio-frequency (RF) accelerators, acceleration in a DLA occurs in the vacuum region of a periodic material structure externally driven by an electromagnetic field. DLA is also in some ways similar to the structure-based wakefield accelerator (SWFA) approach, except that the fields in a DLA are directly excited by a laser rather than through wakefield excitation. 
Due to the approximately 4 orders of magnitude difference in wavelength from the RF to IR regimes, the corresponding feature sizes, cell-to-cell periodicy, and transverse dimensions in a DLA are reduced to the micron to sub-micron scale. The number of accelerating gaps increases accordingly, making a one-kick-per-period approach~\cite {niedermayer:2017} very attractive. 
The similarities between DLA, conventional RF accelerators, and structure-based wakefield accelerators (SWFA), allow similar computational methods and codes to be used. However, since DLA does not rely upon metallic boundaries to confine electromagnetic fields, 
the boundary conditions and laser coupling methods required are in general different. These differences have been found to benefit from computational techniques more commonly used in the design of photonic devices, such as inverse design or adjoint variable method \cite{hughes:avm:2017}. In addition, the smaller size of DLA devices and their use of sub-optical particle bunches requires particle dynamics modeling with high temporal and spatial resolution, and limits the accelerator length amenable to particle-in-cell (PIC) simulation to of order a few centimeters. Thus, transport and particle dynamics over longer (meter to kilometer) scales benefits from the development of more rapid particle tracking codes~\cite {niedermayer:2017}. 

Lastly, novel plasma devices, including capillary discharge plasmas and laser-ionized plasma columns, show promise for application across a range of essential beamline components, including acceleration stages, focusing, energy compensators, and non-destructive beam diagnostics. These systems could enable fundamental advances in high quality electron and positron beams and support TeV-class high luminosity lepton collider concepts. However, modeling these systems with high fidelity requires integrated multi-physics capabilities at computational scales which exceed those of traditional electromagnetic simulation tools. Three-dimensional magnetohydrodynamic or hybrid kinetic solvers provide paths forward to resolving the difficulties associated with these systems.

In the following sections, we report on various key aspects of modeling of AAC systems, such as  high-performance computing, the benefits of ecosystems, sustainability and reliability.

\section{Modeling Advanced Accelerator Concepts}

\subsection{LWFA and PWFA}

The most widely used numerical tool to study plasma accelerators is the particle-in-cell (PIC) algorithm \cite{Birdsalllangdon}, where particle beams and plasmas follow a Lagrangian representation (based on particles trajectories) with electrically charged
macroparticles while electromagnetic fields follow a Eulerian representation (based on field data on a mesh) on (usually Cartesian) grids. Exchanges between macroparticles and field quantities involve interpolations using B-splines at specified orders.
Three-dimensional implementation of the PIC algorithm to model plasma accelerators can be computationally intensive, owing to the disparity of the spatial and time scales in the problem. Laser-driven plasma accelerators are typically (but not always) more computationally intense, compared to beam-driven, since the laser wavelength $\lambda$ must be resolved, and, in an underdense plasma, the laser wavelength is much less than the plasma wavelength $\lambda_p$, which is itself much less than the length of the plasma channel $L_c$, $\lambda \ll \lambda_p \ll L_c$.

Ab initio simulations of plasma accelerators with electromagnetic Particle-In-Cell codes are limited by (i) the number of time steps that are needed to resolve the crossing of the driver through the plasma channel that is orders of magnitude longer, (ii) the number of grid cells (and plasma macroparticles) that are needed to cover the wide disparity of spatial scales that span the physics of the driver beam, the wake and the accelerated witness beam.
Several methods to reduce the computational cost, while maintaining physical fidelity, have been developed.
Reduced dimensionality may be employed in many cases. Indeed,  an ideal plasma accelerator typically has cylindrical symmetry and two-dimensional ($r$-$z$) modeling is often sufficient to describe the relevant physics.   In addition, a truncated series of azimuthal modes may be used to describe nearly cylindrically symmetric cases \cite{LifschitzJCP2009}. 
Furthermore, fluid approximations (moment models) may be employed in situations where plasma kinetic effects are negligible (i.e., when plasma electrons flow are laminar, without crossing of electron trajectories).   

One of the most successful method to speed up simulations is the {\it quasistatic} approximation \cite{Sprangleprl90, Antonsenpop1997}, which relies on separation of time scales to make approximations. The quasistatic approximation decouples the (long time scale) driver evolution and the (short time scale) plasma wave excitation.  This approximation has been particularly successful in modeling PWFAs driven by energetic beams, providing significant computational savings over full PIC (typically three to four orders of magnitude for a multi-GeV acceleration stage). It can also be used to efficiently model LWFAs, provided that self-injection does not occur (this is typically the case for stages operating in linear or mildly nonlinear regimes) \cite{Antonsenpop1997}. 
The massive speedup of the quasistatic approximation comes at the (comparatively small) cost of (i) an algorithm that is specialized and not as widely applicable as the standard PIC method, (ii) approximations that break when plasma electrons accelerate significantly in the wake and get trapped, (iii) more complicated schemes based on pipelines for computing and diagnostics on parallel computers.

For laser-driven plasma accelerators, the wakefield is driven by the ponderomotive force and averaging over the fast laser oscillations in the laser envelope may be employed to implement a ponderomotive guiding center model~\cite{Antonsenpop1997, GordonIEEE2000,Benedettiaac2010,Cowanjcp11,TerzaniPoP2021}, eliminating the need to resolve the fast laser oscillation.  Here the plasma responds to the averaged laser ponderomotive force and may be computed on a coarse grid, while the Maxwell equations determining the laser evolution may be solved on a localized fine grid, resulting in computational savings.  
The speedup offered by using the ponderomotive approximation with a laser envelope description comes at the (comparatively small) cost of a set of equations that is more somewhat complex to solve and parallelize than a standard electromagnetic PIC Maxwell solver.

An alternative to performing a quasistatic approximation to handle the disparity of scales is to shrink the range of scales using the {\it Lorentz-boosted frame method}~\cite{Vayprl07}. With this method, the ultrahigh relativistic velocity of the driver is taken advantage of by using a frame of reference that is moving close to the speed of light in which the range of space and time scales of the key physics parameters is reduced by orders of magnitude, lowering the number of time steps - and thus speeding up the simulations - by the same amount. The tremendous speedup (typically three to four orders of magnitude for a multi-GeV acceleration stage) comes at the (comparatively small) cost of additional complexity for the diagnostics that must reconstruct snapshots of the physics quantities (particles and fields) in the laboratory frame. 

For LWFA and PWFA, the priorities for the modeling of the plasma physics for a single stage will be to (a) demonstrate that reduced models can be used to make accurate predictions for a single GeV-TeV stage (in the linear, mildly nonlinear and nonlinear regimes), (b) demonstrate that the PIC algorithm can systematically reproduce current experimental results (coordinated efforts with experiments) and (c) conduct tolerance studies and role of misalignments. 
There exist many codes that have been used to model plasma accelerators using the abovementioned simulation methods \cite{VayRAST2017}, with new codes emerging regularly.

\subsection{SWFA}
In SWFA, a ``drive" bunch travels through a solid-state structure and excites electromagnetic wakefields. The fields can directly accelerate a trailing ``main" bunch [i.e. in collinear wakefield acceleration (CWA)] or can be extracted to power a separate accelerating structure [i.e. in two-beam acceleration (TBA)]. The simulation of SWFAs is often staged; the structure is first designed to provide the required electromagnetic properties before integrating the beam dynamics and interactions produced by the SWFA accelerator. 

Designing the accelerating structure with the desired electromagnetic modes generally relies on the use of eigenmode solvers to model the structure or one of its unit cells to optimize the field distributions for acceleration while suppressing harmful modes. These tools can be used to efficiently refine the structure geometry independent of the drive beam dynamics. In the case of designing the power extraction and transfer structures (PETS) required in TBA, these solvers are also critical to optimizing the power-extraction coupler. Additionally, time-dependent simulations performed with an FDTD electromagnetic program such as, e.g., {\sc meep}~\cite{meep} or {\sc warp}~\cite{VayCSD12}, over a sufficiently large number of time steps, can provide an understanding of the mode excitation evolution. In such an approach, the beam is modeled by its current distribution. As a first step toward performing integrated simulations, the eigenmode analysis can be used to construct a Green's function. Equivalently, FDTD simulations may be performed with a drive bunch of length much shorter than the mode wavelengths in order to directly extract the Green's function. The obtained Green's functions can be imported into available tracking programs to perform start-to-end simulation of the beam acceleration and guide the design of the SWFA linac (e.g. acceleration staging, understanding the trade-off between accelerating gradient and efficiency). Such an investigation can be performed with programs such as e.g., {\sc elegant}~\cite{elegant} or ~{\sc astra}~\cite{astra}. Very often, they can be limited to the longitudinal beam dynamics using simple 1D-1V models~\cite{litrack,PhysRevAccelBeams.24.051303} to devise the optimum drive-beam shape and the accelerator architecture. The simulation model can be gradually refined by considering the transverse wakefield (modeled with its Green's function approach) to any arbitrary order in complexity via the inclusion of successive multipole terms. This is especially important as most of the advanced structures are fully three-dimensional.

Ultimately, full-electromagnetic simulations coupled with PIC are performed, e.g., for a single stage of the acceleration, to verify the performance of the scheme.  Over the years, several beam-dynamics tracking programs capable of handling collective effects have been altered to include a set of field ``libraries" associated with some SWFA structures~\cite{PhysRevSTAB.15.081304, astra}. This type of model has been used to simulate and guide experiments~\cite{Pacey:2017ynk}. Likewise, a number of PIC codes have been made available with some capability to model SWFAs (compared to L/PWFA, SWFA required matter-type boundaries with specified macroscopic parameters, e.g., electric permittivity, magnetic permeability, or conductivity including their possible dependence on frequency). Finally electromagnetic simulations can also be performed using wakefield models available within the {\sc echo}~\cite{Zagorodnov:206299} and {\sc ace3P}~\cite{ace3p} suite of codes. Ultimately, self-consistent simulation coupling FDTD and PIC algorithm provide a complete understanding of the SWFA process, but their computational expense limits their practicality to the final stages of validation. Improvements to PIC models and performance discussed in following sections, including higher-order Maxwell solvers, moving window, and boosted frame techniques, have been applied with great success to plasma simulations, and could be similarly extended to dielectric structures~\cite{Vayprl07,VincentiCPC2018}. 

Precise tailoring of the wakefield will rely on a high degree of control over the electromagnetic properties of the SWFA so that the development of refined models capable of including frequency-dependent properties of the material are ultimately needed. Likewise, the material-science aspects of the SWFA and the fact that its properties can be altered in the strong-field regime are not captured by any of the programs currently available while recent experiments hinted that the material response to the established fields may ultimately limit the accelerating field~\cite{PhysRevLett.123.134801}.

\subsection{DLA}
The dielectric laser acceleration (DLA) scheme is analogous in many respects to a frequency-scaled version of a conventional accelerator.  Consequently, modelling many aspects of the particle dynamics and transport in a DLA collider can rely largely upon well established accelerator codes and computational methods.  This reduces the need for new code developments or large-scale computing infrastructure to a subset of DLA simulation tasks.  For simulation purposes, a linear collider can be broken into three regimes which require different kinds of modeling:  (1) Injector (source, emittance preparation, and acceleration up to 1 GeV); (2) Accelerator (1 GeV to 30 TeV) and (3) beam delivery, including final focus, crab kissing, IP design, and beamstrahlung.  The proposed injection scheme outlined in Chapter 5, Section 5 of \cite{ALEGRO} is based on superconducting RF technology, and thus is amenable to the same computational methods used to model such systems and will therefore not be addressed here. The main accelerator needs longitudinally coupled structures (traveling wave structures) for energy efficiency and mode stability reasons as described in Chapter 5, Subsection 6.2 of \cite{ALEGRO} . The main accelerator can then be split hierarchically into separately addressed sections corresponding to their respective length scales:
\begin{enumerate}
\item One Optical Period (2$\mu$m)
\item Power coupling cell (10 cm)
\item Focusing cell (20 m at 3 TeV)
\item Betatron period (100 m at 3 TeV)
\item Bunch compression stage (1 km) 
\item Full linac (4.2 km each side at 3 TeV)
\end{enumerate}

An individual DLA cell (one optical period in length) can be straightforwardly simulated by any of a variety of FDTD, FDFD, and FEFD codes combined with various optimization techniques \cite{shin:2013,egenolf:2017,hughes:avm:2017}. The resulting single-cell Fourier coefficients, combined with the corresponding phase and group velocities, can then be combined with 6D tracking codes to obtain long-distance phase space evolutions by symplectic one-kick-per-cell tracking \cite{niedermayer:2017}. The resulting code {\sc{DLAtrack6D}} models the nonlinear fields both accurately and fast, but excluding boundary effects. Beam dynamics based DLA structure synthesis can be done by means of the alternating phase focusing (APF) scheme~\cite{niedermayerPRL2018,niedermayerPRL2020}. Advanced codes and large-scale computing become necessary for the determination of the nonlinear particle trajectories through longer structure segments, since the small feature sizes of the structures need to be resolved. Particle-in-cell (PIC) simulations can be used to model full particle and field dynamics up to a few thousand or tens of thousands of structure periods but becomes computationally prohibitive at longer length scales.  At optical-scale frequencies, a 10 cm interaction distance would be near the limit of what is possible with a modern supercomputer.  A process of using simplified models to construct larger building blocks can be applied to successive levels of the design, using transfer maps to represent larger-scale components such as an entire power coupling cell or focusing cell. 

Moreover, wake functions and beam coupling impedances must be included in the various structures, which can be done both in full 3D or by simplified 2D models in the frequency domain. The most crucial issues here are to find the beam loading and beam break up limits using the longitudinal and transverse wakes~\cite{egenolfPRAB2020}. For the extremely short sub-optical bunches employed in a DLA scenario (with bunch length small compared to transverse size) the theory of beam instabilities might require extension by nonlinear parts of the transverse wakes and transverse position dependent parts of the longitudinal wakes. Such theories have to be validated by extensive PIC simulations. Moreover, the consequences of slight steering errors in the 10-nanometer range, leading to severe average beam power deposition in the structures needs to be studied. 
While much of the above simulation work can be accomplished by use of existing codes and/or well-established computational techniques, the beam dynamics in the final focus of a DLA collider needs to be completely redeveloped for DLA-type beams, unless, as assumed in Chapter 5, Section 5 of \cite{ALEGRO}, the microbunching is washed out prior to interaction at the IP.  Direct collision of trains of extremely short, low intensity bunches with high repetition rate would behave very differently than conventional bunch crossings when it comes to beamstrahlung or crab kissing.  Fast beam-beam interaction codes are available, but have yet to be adapted to attosecond bunches.


\subsection{Structured plasmas}

A structured plasma is a plasma system whose density, temperature, composition, and ionization state are tailored to enhance a desired interaction between a beam or laser with that plasma system. Technologies leveraging structured plasma are increasingly central to advanced accelerator concept schemes. Controlled plasma channels are integral to increasing the peak energy and quality of laser-accelerated electron beams~\cite{Geddes_2004, Leemans_2006, Ibbotson_2010}. For LPA schemes, generation of a plasma channel with specific radial density profiles enable matching of the drive laser to the plasma, reducing diffraction while maintaining peak on-axis intensities~\cite{Sprangleprl90}. Longitudinal density control can be used to trigger injection via density downramp~\cite{Suk_2001}, and to modify dephasing through longitudinal tapering~\cite{Sprangle_2000}. Similarly, PWFA schemes have benefited from the generation of meter-scale plasma channels with near-uniform density ~\cite{Blue_2003, Blumenfeld_2007, Litos_2014}. The use of pre-ionized channels reduces head erosion caused by defocusing of the drive beam, thereby permitting longer acceleration lengths~\cite{Hogan_2010, Green_2014}. Finally, recent demonstrations of positron wakefield acceleration have relied on pre-ionized hollow-channel plasmas~\cite{Gessner_2016}. Hollow-channel plasmas improve accelerating gradients and phase-stability compared with uniform plasma channels~\cite{Lee_2001}, and may enable the independent control of focusing and accelerating fields for emittance preservation~\cite{Schroeder_2013}. Finite radius plasma columns have also been proposed as plasma targets to accelerate positrons in an electron driven PWFA~\cite{Diederichs_2019}.

Beyond sources and stages, structured plasmas have found application as flexible focusing elements. Discharge capillaries are capable of producing orders-of-magnitude larger magnetic field gradients than traditional quadrupoles or solenoids~\cite{vanTilborg_2015}, subsequently enabling the compact staging of plasma accelerators~\cite{Steinke_2016}. Alternative approaches include passive plasma lenses, consisting of a narrow plasma jet outflow generated by laser pre-ionization~\cite{Lehe_2014,Doss_2019}, which can provide comparable focusing with sufficiently high density. Structured plasmas have been employed as tunable dechirpers, capable of removing correlated energy spreads from GeV-scale electron beams~\cite{DArcy_2019}. Finally, a promising class of non-destructive diagnostics with high spatiotemporal resolution will rely on controllable plasma densities under interaction with electron and laser beams~\cite{Scherkl_2019}.

Modeling requirements for structured plasma systems vary significantly with the device type, scale, and application. For capillaries, proper modeling requires the characterization of the discharge current and resulting plasma transport properties, including electrical resistivity and thermal conductivity. Magnetohydrodynamic (MHD) codes are well suited to capturing the basic physics of these systems, while maintaining larger timesteps and reduced resolution requirements from kinetic approaches. Significant progress has been made in demonstrating their agreement with 1D analytical models and experimental results for waveguides and active plasma lenses~\cite{Bagdasarov_2017,Gonsalves_2019}. The coupling of such MHD codes with laser envelope models have contributed to record-breaking LPA acceleration of electrons to 8 GeV in 20 cm capillary \cite{Bagdasarov_2021}. Similarly, active plasma lens studies have benefited from the application of MHD to reproduce species-dependent nonlinearities in current flow and magnetic field~\cite{vanTilborg_2017,Lindstrom_2018,Cook_2020}.

Pre-ionized plasma sources, of the kind implemented for PWFA stages, passive plasma lenses, and hollow-channel plasmas, have traditionally leveraged high pressure gas jets or plasma ovens to produce the necessary neutral density profiles. Capturing the gas channel or sheet characteristics may necessitate hydrodynamic simulation on $\mu$s-timescales, while the subsequent laser ionization requires the computation of laser propagation and self-consistent field ionization profiles on fs-timescales. Furthermore, the pre-ionized plasma does not constitute a local thermal equilibrium (LTE), therefore the LTE dynamics implemented by most MHD codes is insufficient to capture the ionization and heating dynamics. Moreover, vacuum-plasma interfaces are of interest for matching incident drive beams~\cite{Xu_2016}, and multi-species plasmas have been employed for high-brightness injection scheme~\cite{Deng_2019}. In these cases, kinetic codes, for example particle-in-cell techniques, may be coupled with hydrodynamic codes to obtain reasonable approximations, or alternatively, first principles models may be employed~\cite{Xi_2013,Lehe_2014,Manahan_2019,Doss_2019,Gessner_2020}.

\subsection{From ultra-accurate to ultrafast}

The development of high-gradient particle accelerators demands for predictive modeling tools that range from 
(i) very detailed, full physics and dimensionality, first-principle kinetic simulation tools that are needed for detailed runs for physics studies (typically based on the Particle-In-Cell method)~\cite{LOI_eva},
to 
(ii) ultrafast simulation tools  that are needed for ensemble runs 
for design studies (using a combination of, e.g., reduced physics, 
low dimensionality, low resolution, artificial intelligence 
(AI)/machine learning (ML) surrogates~\cite{LOI_ML, LOI_ML2}).

\subsubsection{End-to-end Virtual Accelerators (EVA)}
The realization of software that are capable of end-to-end virtual accelerator (EVA) modeling (or ``virtual twins'') has been identified as a Grand Challenge of Accelerator and Beam Physics \cite{LOI_ABPRoadmap}.

Thorough modeling of the full particle accelerator in individual parts
(injector, magnets, beam dynamics, etc.) is already an important aspect 
of particle accelerator development; one, without which no project 
nowadays can proceed to the building stage. However, simulating the 
system in many small units poses two significant challenges: 
(i) The various codes used for the individual parts are often from different
eras, written in different languages (C/C++, Fortran, Python, etc.),
and use different standards for data  I/O (particles, fields, etc.), 
slowing down the design and simulation process; 
(ii) Without multiphysics couplings, subtle, but important effects 
(collective effects, halo, coherent synchrotron radiation, etc.) might not
be considered in the design, ultimately leading to issues during 
commissioning and the need to modify the machine (shimming, 
additional steering, shielding, etc.) and longer downtimes due to added
radiation. Furthermore, opportunities for more efficient working points that can be achieved only from system optimization of the entire accelerator may
be missed entirely.

Thus, there is a clear need for full physics 6-D computer simulations 
of the entire accelerator system that incorporate all components (including both conventional and AAC sections), all pertinent physical 
effects, and that execute fast and reliably. Furthermore, these EVAs
should be able to leverage modern computing infrastructure like HPC clusters and GPU computing, and fully integrate AI/ML tools to maximize efficiency 
for practical applications. 
The development of such tools requires continuous advances in fundamental beam theory and applied mathematics, improvements in mathematical formulations and algorithms, and their optimized implementation on the latest computer architectures. 


\subsubsection{Optimization \& Ultrafast AI/ML surrogate models}
Surrogate models can be built by training a neural network on a smaller (compared to 
other optimization techniques) set of high fidelity simulations (often particle-in-cell)
that coarsely maps out the hyperspace of possible input parameters. Non-simulated sets of input parameters can then be approximated by the surrogate model. 
This can be highly useful for optimization and online feedback about the accelerator.
Some examples of successful use of surrogate models in particle accelerator optimization are 
\cite{adelmann:surrogate1, van_der_veken:ml1, edelen:ml1, edelen:ml2}.
A speedup of one to several orders of magnitude compared to 
conventional techniques has been observed in these cases.

Other examples, specifically using Bayesian Optimization using Gaussian Process models, 
are the tuning of SwissFEL \cite{kirschner:bayesian1, kirschner:bayesian2} and 
the Linac Coherent Light Source (LCLS) \cite{duris:bayesian1}.

These surrogate models will have strong impact on accelerator modelling and 
optimization through the ultrafast execution of the trained models.
They can be trained for full start-to-end virtual accelerators or individual
components. The ultrafast execution enables quick turnarounds for design
optimization and as an on-line feedback tool during commissioning, tuning,
and operation of an accelerator system.

Important steps forward to make surrogate models widely usable have been identified 
in Ref. \cite{LOI_ML}. The main paths forward are \emph{finding the best
practices} for various different accelerator systems and modeling needs thereof, and 
\emph{developing a robust and flexible framework} for surrogate modeling.
\section{High-Performance Computing}

High-Performance Computing (HPC) continues to be a tightly connected domain for advanced accelerator modeling research.
One of the main reasons for this is the large computational model size for full-fidelity PIC simulations, especially when describing kinetic effects in relativistic laser-matter interaction.

Over the last decade, HPC machines established so-called hardware accelerator components, most notably general purpose programmable graphics processing units (GPUs), in a majority of leadership-class machines.
Such novel compute hardware provides an order of magnitude higher computational density compared to traditional computing hardware under reduced energy costs.
GPUs are highly parallel computing devices and need algorithms that are 10- to 100-thousandfold parallelized compared the the conventional parallelism in CPUs (10- to 100-fold parallel).
When redesigning algorithms for modern compute architectures, PIC codes for accelerator modeling can usually achieve an order-of-magnitude speedup for time-to-solution \cite{PIConGPU2013,Myers2021}.

\subsection{Performance, portability and scalability}

\subsubsection{Performance}
Performance increase for modeling software is generally desirable, even outside of large-scale HPC runs, and improves overall scientific productivity when studying particle accelerators at all scales.
For instance, most optimization and machine-learning workflows are GPU-accelerated as well to increase computational throughput for studies, while ultraprecise numerical schemes can often use GPU-accelerated linear algebra and FFT routines; thus GPU acceleration also benefits small model sizes.

Developing software for multiple GPU-platforms in parallel to existing CPU architectures is an undertaking that requires the redesign of many existing algorithms and code-bases.
Programming models and languages evolve rapidly and the need to port algorithms to significantly more fine-grained parallelism is a good incentive to rewrite and adopt modern software practices and popular programming languages.


\subsubsection{Portability}
Portability of implementations is important as many scientists work with three major platforms (Linux, macOS, Windows), although the former two (Unix-like) variants are certainly dominating with HPC developers and machines.
An additional challenge arises with the diversity in GPU and other compute hardware, which ideally should be addressed with \textit{performance-portable} software design.
In such a design, an algorithm is ideally implemented only in a single, abstract form and then specialized/optimized for the specific hardware needs at compilation-time.

Performance portability relies on using and contributing to standard, industry-supported programming languages to allow such a single-source approach.
Performance portability layers (e.g., Kokkos~\cite{kokkos}, Alpaka~\cite{Alpaka}, Raja~\cite{Raja}) are currently implemented predominantly as modern C++ libraries with other languages lacking significantly in release time, compiler variety and robustness.
Common numerical and data management libraries (e.g., AMReX~\cite{AMReX}, CoPA~\cite{Copa}) build on such performance portability layers and only as the next steps, domain-specific libraries and applications are implemented to benefit from all aforementioned developments.

\subsubsection{Parallel scalability}
Parallel scalability on leadership-scale machines continues to be only achievable with additional parallelization over multiple coupled compute nodes (each powered with CPUs or GPUs).
Methods that (a) split the computational domain into local sub-problems, and (b) ideally overlap communications (usually of data in guard cell regions) between collaborating nodes with computation within each node, are essential for such capability simulation runs.
Furthermore, at large-scale, load balancing of particle-based methods as well as refinement of fidelity needs (e.g., adaptive mesh-refinement) of the simulation domain become important to make optimal use of available resources.


Looking at the rapidly evolving HPC landscape, many themes in advancing accelerator and beam modeling software are common, e.g. Poisson solvers, robust particle pushers, need to model hybrid particle accelerators combining conventional and advanced accelerator elements, etc.
It is thus evident that collaborative efforts between conventional and advanced accelerator development can benefit the community, relying for instance on similar or compatible low-level parallelization layers, but also coordination when implementing and sharing domain-specific numerics can ensure that scalable solution can be readily extended for research needs.

\subsection{Advanced algorithms}

In addition to developing codes that use the latest architectures as efficiently as possible, it is essential to constantly reassess the capabilities and limitations of existing algorithms, and develop new ones that improve stability, accuracy and/or performance. The extreme computational intensity required for full PIC modeling of LWFA has fostered many innovations in Particle-In-Cell methodology. In particular, the impetus to complete these simulations in a reasonable time using a Lorentz boosted frame has led to active research to mitigate some numerical instabilities that are otherwise particularly violent.

In a Lorentz boosted frame PIC simulation, the plasma column, which is at rest in the laboratory, is moving near the speed of light in the frame of the simulation. This results in coupling between the fields on the (static) computational grid and the plasma modes and its aliases that lead to resonances and the so-called {\it numerical Cherenkov Instability} \cite{GodfreyJCP1974,GodfreyJCP1975}. Research that was spurred by the drive to perform Lorentz boosted frame LWFA simulations has led to tremendous progress in the understanding of the instability and its mitigation \cite{Vaypac09,Martinscpc10,VayJCP2011,VayPOPL2011,GodfreyJCP2013,XuCPC2013,GodfreyJCP2014,GodfreyJCP2014b,GodfreyIEEE2014,GodfreyCPC2015,YuCPC2015,YuCPC2015-Circ,LehePRE2016,KirchenPoP2016,KirchenPRE2020,FeiCPC2017,PukhovJCP2020}.
One of the most notable of these advances is the {\it Galilean Boosted Frame approach}, which consists in using the ability to integrate Maxwell's equation analytically in time over one time step of the {\it Pseudo-Spectral Analytical Time-Domain} (PSATD) Maxwell solver \cite{Habericnsp73,VayJCP2013} to synchronize all the fields and plasma modes with a Galilean frame that follows the plasma in the boosted frame, hereby removing the resonances between all the corresponding modes and leading to stability \cite{LehePRE2016,KirchenPoP2016}. 

This research has also spurred advances in the PSATD solver itself, enabling its efficient usage in parallel using domain decomposition (where one FFT is performed per MPI subdomain) with arbitrary order of accuracy (and locality) of the spatial derivatives \cite{VayJCP2013,VincentiCPC2016,JalasPoP2017,VincentiCPC2018,KirchenPRE2020}.
This research is ongoing and very active. Most recently, a novel algorithm was proposed that takes again advantage of the PSATD ability to integrate analytical the field over a time step to average the field that is used to push particles over one time step \cite{ShapovalARXIV2021}. This overcomes a longstanding barrier of boosted frame simulation that were limited to small time steps by the Courant-Friedrich-Lewy (CFL) condition, when the transverse cell size of the grid $\Delta x$ is smaller than the longitudinal cell size $\Delta z$, $c \Delta t \leq \Delta x< \Delta z$, where $c$ is the speed of light in vacuum. With the novel algorithm, the simulation is still stable even when $c \Delta t> \Delta x < \Delta z$. In another recent advance, a novel {\it hybrid} PSATD PIC scheme was proposed that combines the advantages of stability of nodal PIC methods and efficiency of staggered PIC methods, demonstrating speedups of boosted frame full PIC LWFA and PWFA simulations by factors of two \cite{ZoniARXIV2021}. 

In addition to reducing the number of time steps, techniques to reduce the number grid cells that are needed are being developed. Mesh refinement, a technique that is well established in the modeling of fluids, has been used with success in the modeling of conventional accelerators with electrostatic PIC solvers \cite{Vay2004}, and has been explored more recently for the modeling of plasma accelerators with electromagnetic PIC solvers \cite{VayCPC2004,VayCSD2012,WarpXEAAC2017,WarpXAAC2018}. 
Reduction  of dimensionality using 2D planar or axisymmetric geometries is a very common way to reduce the computational workload. Fourier decomposition in azimuthal modes, a technique that has been borrowed from other areas (including conventional accelerators), has been introduced in this community in 2009 \cite{LifschitzJCP2009} and been adopted by more codes since, using full PIC with boosted frame \cite{Lehe2016}  or quasistatic PIC \cite{LiCPC2021}. The rapid adoption by the community of the code FBPIC \cite{FBPIC} demonstrates the effectiveness of combining the full PIC approach with reduced dimensionality using azimuthal Fourier decomposition in a software that is easy to install (as a Python module) and yet efficient on both CPUs and GPUs.

It is important to note that the constant development of novel algorithms, which has been fostered by the extraordinary numerical and computational challenges associated with the modeling of high-gradient accelerator components, has found application to other related fields. 
For example, the development of distributed parallel PSATD PIC \cite{VayJCP2013,VincentiCPC2016,VincentiCPC2018,Kallala2019} has spurred research into the numerical properties of Maxwell solvers when modeling plasma mirrors with electromagnetic PIC codes.
These types of simulations were shown to benefit greatly from ultrahigh-order pseudo-spectral solvers of order 50 or more \cite{Blaclard2017}, leading to simulations of plasma mirror experiments and applications \cite{Leblanc2017,Vincenti2019,Chopineau2019,FedeliPRL2021} that are otherwise out of reach to standard PIC methods with finite-difference Maxwell solvers. Another example is the development of the novel hybrid PSATD scheme that was mentioned above, which has found application in the modeling of electron-positron pair creation in PIC simulations of high-field physics \cite{ZoniARXIV2021}, with speedups by an order of magnitude or more over conventional methods, exemplifying the cross-fertilization of the progress in new algorithms across applications.

\subsection{Scalable I/O and In-Situ Analysis}


While modern programming models, software design and advanced algorithms ensure that simulations can execute efficiently, one further challenge emerges to scientific productivity on HPC hardware:
When comparing the computational peak performance evolution to the parallel system storage bandwidth of HPC machines as in Table \ref{table:olcf}, the gap between potential simulation fidelity that can be computed and data that can be stored for analysis widens with every new machine generation.

\begin{table}[h!]
 \centering
 \begin{tabular}{||c c c c||} 
 \hline
 System Specs & Titan & Summit & Frontier \\ [0.5ex] 
 \hline\hline
 Peak Performance & 27 PFLOP & 200 PFLOP & >1.5 EFLOP \\ 
 \hline
 Storage Bandwidth & 1 TB/s & 2.5 TB/s & 2-4x Summit \\
 \hline
 Ratio & 27:1 & 80:1 & 150-300:1 \\
 \hline
 \end{tabular}
 \caption{Evolution of leadership-scale machines at the Oak Ridge Leadership Computing Facility with respect to peak performance and storage bandwidth, according to \href{https://web.archive.org/web/20210715000401/https://www.olcf.ornl.gov/frontier/}{https://www.olcf.ornl.gov/frontier/}.}
 \label{table:olcf}
\end{table}

Following this trend, modeling applications that run at a system's capability and follow the traditional workflow of simulation + post-processing will spend more of their time in large-scale data input and output, which eventually can dominate the time-to-solution.
Driven by this trend, several steps can be taken by simulation developers.

As a first step, adopting scalable, parallel I/O libraries with adequate data layouts \cite{Wan2021} can extent the applicability to truly make optimal use of parallel filesystems \cite{Huebl2017}.
Implementing these libraries in both data producing as well as data consuming codes opens another possibility to standardize on and share the community solutions, as demonstrated in the openPMD project \cite{openPMD}.

As a second step and long-term solution, data analysis and processing need to be transitioned to in situ-processing, running online with the modeling simulation and avoiding traditional, file-based long-term storage for raw data.
This approach essentially moves the design of data analysis to the setup and design phase of a simulation that, similar to an experimental setup, needs to plan the locations, acceptance and dynamic ranges of ``virtual diagnostics'' before even starting a high-fidelity modeling run.
Virtual diagnostics can reduce data sizes and thus data throughput needs by orders of magnitudes, e.g., by calculating beam momenta, phase space histograms, spectra and even 3D videos as a simulation runs instead of storing realistically trillions of particles in regular intervals to disk for later analysis.
While in-situ data processing and analysis are not new, and have even been mainstream in some accelerator codes or frameworks (e.g., Warp~\cite{VayCSD12}, PIConGPU~\cite{PIConGPU2013}), the discrepancy between peak performance and storage bandwidth reported in Table \ref{table:olcf} triggers the development of community libraries that enable a more systematic description of new diagnostics in an in-situ approach, ensuring high efficiency and scalability.

A challenge arises in the rapid setup of in situ processing pipelines, since typical approaches of implementing a virtual detector in the simulation code itself can be more complicated compared to scripted (serial) analysis workflows on files.
Innovative software design can potentially alleviate this challenge, with approaches such as streaming data with close-to-file-like analysis syntax \cite{Poeschel2021} or embedding of interpreters into running simulations to execute scripts on in-memory data.
Continued research in this area and close collaboration with computer science and engineering teams will be essential to ensure that flexible, science-case specific analysis can be designed by accelerator and beam physicists.

\section{Community ecosystem}

Simulations are critical to the design, development and operation of all accelerator facilities, some costing in the billions of U.S. Dollars. 
The simulation of all the components and physical processes that are involved is a very wide-ranging and complex endeavor.
While a lot of efforts have been aimed at the development and maintenance of very useful and successful simulation tools, they were largely uncoordinated. 
This resulted in a rich collection of (often relatively small, with some large and sophisticated) codes that are not interoperable, use different I/O formats and quite often duplicate some physics functionalities using the exact same underlying algorithms (e.g., many codes have standard Particle-in-Cell cores that are distinct but algorithmically identical), impoverishing the effective diversity of the whole.

Accelerator and beam physics software, like many other types of software, are subject to a `natural cycle', driven by, e.g., developers retiring or moving to other projects, evolution of programming languages or computer hardware, leading to regular retirement or loss of codes that may need to be rewritten from scratch. The rewriting can be particularly costly when the previous code is poorly documented (if at all) and the original developer(s) is/are not available.

\begin{figure}[tb]
    \centering
    \includegraphics[width=6in]{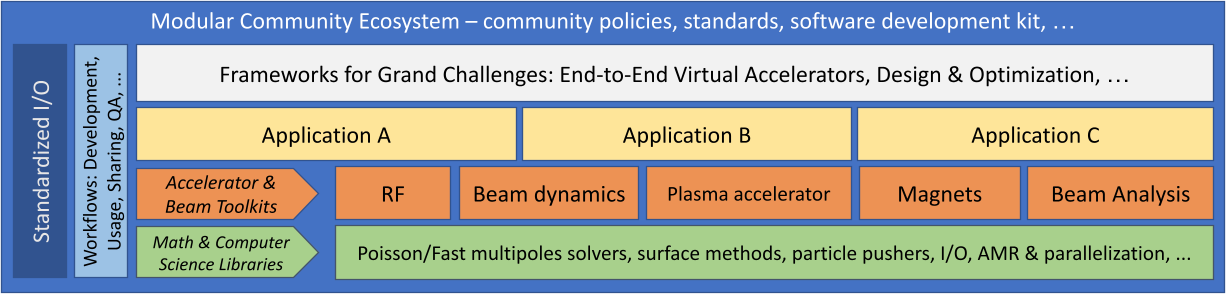}
    \caption{Diagram of a possible {\it Ecosystem} for particle beam and accelerator modeling.}
    \label{f:ecosystem}
\end{figure}

A concrete community ecosystem for accelerator and beam modelling could be structured as in Figure~\ref{f:ecosystem}: individual components are represented as boxes with dependencies building on top of each other.
The fundamental building blocks are implementations of solvers/numerical schemes, efficient I/O, parallelization and a performance portability layer (\textit{Math \& Computer Science Libraries}).
Depending on this functionality, domain-specific libraries for, e.g., RF, beam or plasma modeling can be implemented (\textit{Accelerator \& Beam Toolkits}) that then act as toolkits to be combined into concrete \textit{Application}s.
Compatibility as well as development productivity, usage, and quality assurance of an ecosystem would be coordinated with standardization (e.g., for I/O and data layouts) as well as common workflows, e.g., shared continuous benchmarks and practices.
Overarching frameworks then implement concrete workflows that combine applications or toolkit components in an innovative way for the study of complex, integrated research questions.

\subsection{Integrated workflows}
Integrated workflows are often needed to answer complex contemporary research questions, such as the start-to-end description of a hybrid accelerator with conventional and plasma elements.
Furthermore, all optimization, AI/ML training as well as exploratory studies benefit from establishing and maintenance of workflows.
These workflows might (a) link several codes together to solve a larger problem or solve a problem with more physical processes, (b) benchmark and validate the various codes against each other and known solutions, and (c) establish reproducibility for new scientific results.

\subsubsection{Standardization of input \& output}

At the moment, many simulation codes are developed independently with little coordination between various groups.
In order to establish workflows that span multiple applications, a typical approach so far is to rely on file-based data exchange and code-specific high-level wrapping of input options.
Along these lines, the Consortium for Advanced Modeling of Particle Accelerators (CAMPA~\cite{CAMPA}) supports (among other activities) the development of standardized input and output formats through the openPMD~\cite{openPMD} and PICMI~\cite{PICMI} projects.

The Open Standard for Particle-Mesh Data (openPMD) is a meta-data standard for research data, adding scientific self-description on top of modern, portable, scalable file-formats such as ADIOS~\cite{ADIOS2} and HDF5~\cite{HDF5}.
openPMD achieves this by collaboratively defining an open standard document that evolves in versions, similar to software.
Community members then implement openPMD data handling in their simulation codes and analysis types either directly against supported file formats or use an open source library layer, openPMD-api~\cite{openPMDapi}.

The standard input format for Particle-In-Cell codes (PICMI) strives to unify the usage of accelerator modeling codes by standardization on an API for simulation design.
Conceptually, PICMI defines a common, explicit PIC modeling interface (API) for a wide range of numerical and initialization options.
A PICMI script can be run with multiple codes, given that both implement the requested features, and simplifies comparability of physics cases tremendously.
Code "backend" choices can be picked based on performance or numerical needs, without breaking context for the user and relearning the simulation design for reach specific application.

\begin{figure}[tb]
    \centering
    \includegraphics[width=6in]{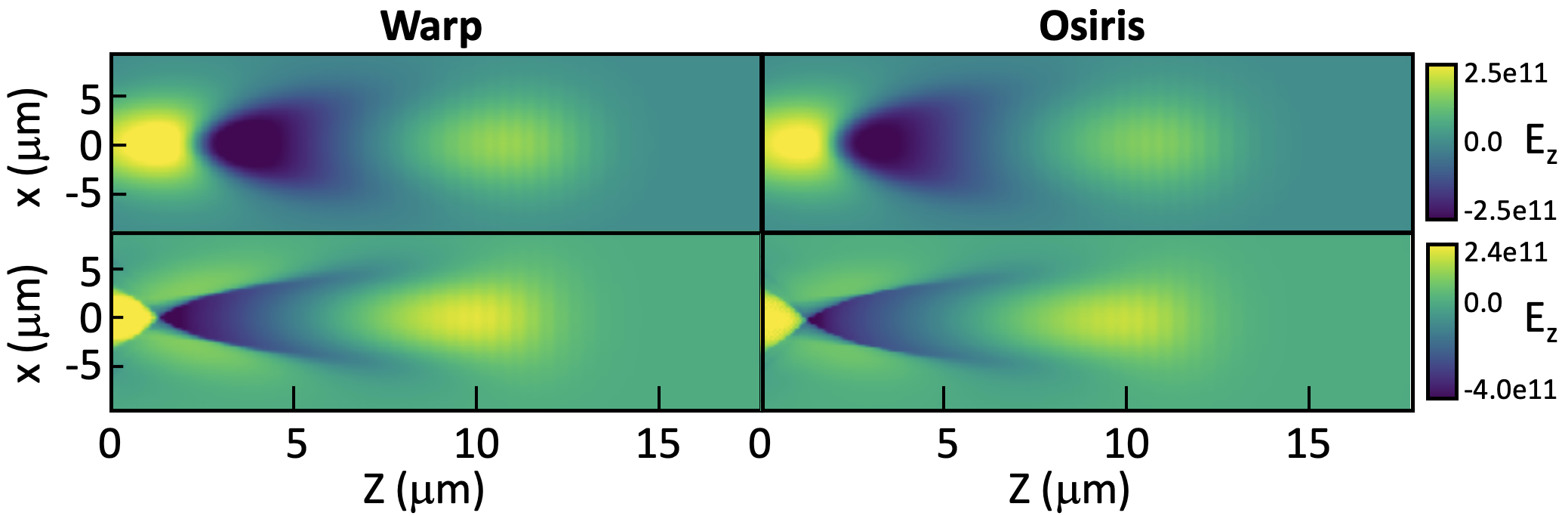}
    \caption{Longitudinal electric field (in V/m) in a laser-driven plasma acceleration stage at two times (top: $t\approx 300$fs, bottom: $t\approx 600$fs) along the laser propagation
from 2-D PIC simulations with: (left) Warp; (right) Osiris. Plots are based on rendering from the openPMD-viewer.}
    \label{f:warp_osiris_openpmd}
\end{figure}

As part of the CAMPA efforts to foster compatibility and coordination between particle accelerator modeling applications, a code-validation workflow between the Warp~\cite{VayCSD12} and Osiris~\cite{Osiris} code has been carried out.
In the study on the two codes in Figure~\ref{f:warp_osiris_openpmd}, both independent code bases modeled the same physical setup of an LWFA (with a laser driving a wake in a plasma column) and used standardized openPMD output and the same, community-developed post-processing tool (openPMD-viewer) to analyze the data.
A one-to-one comparison to machine precision between the two codes was beyond the scope of the study because: (a) while using the same physical parameters to initialize the simulations with each code, some secondary parameters were left unspecified such as, e.g., the exact phase of the laser oscillations, (b) some numerical aspects are different such as, e.g., the method of initialization of the laser in the boosted frame.
Nonetheless, it is rewarding to observe on Figure~\ref{f:warp_osiris_openpmd} that the two codes predict very similar wake evolutions and dephasing of the laser as it propagates through the plasma column.

Another series of comparisons is underway that also incorporates a common Python input script using PICMI, and including more codes, in the workflow, to validate the workflow concept for further adoption and development.


\subsection{Integrated frameworks}
Integrated developments of frameworks could implement more tightly coupled workflows on a library level instead of relying on high-level application coupling.
As presented at the beginning of the section in the concept of an modular software ecosystem in figure~\ref{f:ecosystem}, modular, cross-cutting software designs can provide reusability, efficient development investments, and compatibility.
Built-in features of lower-level software such as GPU-support, multi-node parallelism, translation to AI/ML frameworks and multi-physics support can, if evolved together, be inherited by a wide community.
In particular, common or compatible in-memory data structures have not been attempted in accelerator and beam modeling so far, but would provide significant performance and software maintenance benefits for tightly-coupled, co-developed physics modules for integrated modeling frameworks that address grand challenges.

Several community examples exist to demonstrate modular development.
For instance, the codes WarpX~\cite{VayPoP2021} and PIConGPU~\cite{PIConGPU2013} as well as several analysis frameworks share the development of the I/O library openPMD-api~\cite{openPMDapi}, which provides access to high-performant, portable low-level I/O libraries.
WarpX, FBPIC implement PICMI assisted by a shared, central interface library, with implementation underway in other codes (e.g., OSIRIS).
WarpX, HiPACE++~\cite{HipacePP} and related codes share the data structures from AMReX~\cite{AMReX} and parts of their code bases.


\subsubsection{Community development}

This transition to a more coordinated and collaborative approach is also driven by the need to transition a large body of software from CPUs to GPUs, a transition that is more disruptive than most past transitions, including the transition from serial to parallel codes (and it is anticipated to be one of many transitions needed to adopt to the rapidly evolving computer hardware).
Three dimensional simulations of advanced accelerator concepts have been around since the early 2000's (for wakefield accelerators).
Ever since then, more realisms (e.g., ionization, QED effects), better numerics, and advanced algorithms have been added to almost all code bases, and code developers have developed best practices for adding new modules while maintaining the performance needed to run on state-of-the-art HPC centers around the world. 
Thus, instead of porting solely to GPUs, accelerator modeling teams already teamed up with computer scientists and embrace libraries that abstract computer hardware specific details for performance and sustainability.
Establishing more community standards, funding reliable software dependencies and reusable physics modules are the logical steps to speed up development of code bases further.


An efficient mean for guiding community development is the collaborative setting of and adherence to common policies.
Rather than starting from scratch, one can adopt an existing libraries and codes and build a beam and accelerator modeling ecosystem on it.
As a blueprint, one could build on community development concepts of the Interoperable Design of Extreme-scale Application Software (IDEAS) project \cite{IDEAS} and its Extreme-scale Scientific Software Development Kit (xSDK) \cite{xSDK}.

The main goal of the IDEAS project is to ``help move scientific software development toward an approach of building new applications as a composition of reusable, robust, and scalable software components and libraries, using the best available practices and tools.''\cite{IDEAS}, while the 
xSDK project is being developed to ``provide a coordinated infrastructure for independent mathematical libraries to support the productive and efficient development of high-quality applications'' \cite{xSDK}:

\begin{quote}
\small
Rapid, efficient production of high-quality, sustainable extreme-scale scientific applications is best accomplished using a rich ecosystem of state-of-the art reusable libraries, tools, lightweight frameworks, and defined software methodologies, developed by a community of scientists who are striving to identify, adapt, and adopt best practices in software engineering. The vision of the xSDK is to provide infrastructure for and interoperability of a collection of related and complementary software elements—developed by diverse, independent teams throughout the high-performance computing (HPC) community—that provide the building blocks, tools, models, processes, and related artifacts for rapid and efficient development of high-quality applications.
\end{quote}

Although xSDK targets ``extreme-scale scientific applications'' of relevance to exascale supercomputing, its derived community policies apply well to all scales of computing, from laptops to clusters or Cloud computing.  Hence, the paradigm and tools are readily applicable to the full set of modeling needs of the particle accelerator community.

Individual packages in such an ecosystem are composable libraries, frameworks, and domain components developed by individual groups in the community.
Each package publishes technical design documents, documentation (reference, tutorials, how-to guides), API definitions, and a concrete implementation \cite{LOI_industry}.  In order to solve a specific physics case, a typical application uses functionalities from several packages, often in an innovative way.  

There are many advantages to adopting policies and tools such as the IDEAS/xSDK:
\begin{itemize}
   \item Community policies have been established over years by teams of specialists in scientific software development and computing, and they include best practices in software development.
    \item For reusable components, the policies require the use of {\it permissive open source licenses} (``Non-critical optional dependencies can use any OSI-approved license.'') that allow reuse of source and binary code, including for commercial and proprietary applications, fostering collaborations across laboratories, academia and industrial partners.
   \item A wide-ranging set of open source, interoperable tools from a variety of backgrounds (including numerical solvers, e.g., Hypre, multi-parameter optimizer, e.g. LibEnsemble, and mesh-refinement frameworks, e.g. AMReX) can be combined and used as foundation of accelerator and beam physics toolkits and codes.
    \item A wide community of developers that can help improve software capability and sustainability across projects.
    \item Partial overlap in functionality is not problematic and in fact improves the diversity of the whole.
    \item Time of development and maintenance of domain specific software is greatly reduced, as the domain scientists can concentrate on the domain-specific functionalities while lower-level, cross-cutting, numerical packages are maintained by dedicated specialists.
    \item Portability across platforms (CPUs, GPUs, etc.) of low-level libraries ensures portability of the software that is built upon them.  Building new tools on these low-level libraries greatly reduces the overall burden on the community for porting codes to new platforms.
    \item The domain-specific packages that are built, following established policies are portable across platforms and interoperable, and can  be further combined to form larger toolkits and codes.
    \item Since packages are reused across software and duplication is minimized, bugs and inefficiencies are spotted earlier.
    \item Many codes, libraries and packages exist in the community that can be reused as building blocks, and progressively modernized and blended in a module that is portable and runs efficiently on modern CPUs and GPUs. Hence, not all functionalities have to be rewritten from scratch at once, offering a progressive path from the current status to an ecosystem of accelerator modeling tools. 

\end{itemize}

It is to be emphasized that the goal is not to propose that only one code be developed to study, e.g., RF accelerators or plasma-based accelerators. 
On the contrary,
the goal is to provide a coordinated infrastructure that enables inclusive and collaborative development of modeling tools for the community, and for these tools, to follow modern practices for scientific software: be portable, efficient, robust and leading to consistently accurate and reproducible results.

\section{Sustainability \& Reliability}

An important aspect of the development of scientific modeling tools is to ensure that the corresponding software is robust, easy to use and can be extended for new science cases. This is made more difficult by the fact that these tools are oftentimes constantly evolving, with new capabilities being continuously added to a given software. Here we mention a few important practices \cite{LOI_industry} that facilitate this process.

\subsection{Code robustness}

Because modeling tools are being continuously improved and extended by developers, there is always a risk that a given change to the source code may introduce bugs and produce erroneous results in some specific cases. It is therefore paramount that the developers track the changes to the source code (with a version control tool such as \texttt{git}), and that they simultaneously maintain a comprehensive suite of tests that the code should always successfully pass. In the case of scientific modeling codes, these tests can consist of a number of physical setups for which there are well-known theoretical predictions. It can thus be checked that the results of the code conform to these predictions. In order for these tests to be most effective, they need to be run \emph{automatically} whenever the source code is modified -- a process known as \emph{continuous integration}.

In addition, the development of large-scale scientific tools nowadays involve a whole community of developers and users rather than a single, isolated developer. Therefore, the robustness and quality of the software is generally greatly improved by the adoption of collaborative development platforms. These online platforms allow the developers to easily and efficiently review each other's changes to the source code, before these changes are incorporated in the mainline version of the code. They also provide a central venue for users to raise issues and questions, and generally communicate their needs with the developers.

Finally, another aspect of a code's robustness is its interface with the user. As the code evolves, the interface with the user often needs to be modified so as to enable functionalities or generalize existing ones. It is not uncommon that these changes break some of the users' scientific workflows. Thus, it is important for the developers to regularly publish releases of the source code, where these changes are clearly documented. The different releases of a given software can be made citable and available through an online archival service, so that users can roll back to a previous release if needed and scrutinize changes in a reproducible manner.

\subsection{Usability}

In order to maximize the impact of a given software tool on its scientific community, it is also key that this tool be user-friendly. For example, a common obstacle to the adoption of some scientific tools is their complex installation procedure, especially if many dependencies are involved. 
The developers can drastically simplify the installation procedure of a given software tool by making it available through modern package managers \cite{spack, conda, pip}. These package managers can automatically download and compile the source code as well as its set of dependencies. They can also ensure that compatible versions of the dependencies are being used. Importantly, some of these package managers can allow users to have the exact same installation workflow on large-scale HPC resources as on their desktop workstation.

Another key aspect of the user experience is the code documentation. For scientific modeling tools, the documentation usually includes sections on how to install and use the code, as well as a description of algorithm being used along with relevant references. Since, again, scientific tools are often continuously evolving, it can be beneficial that the documentation be generated automatically, whenever the source code is modified or whenever a new release is published.
Just as the modeling tools themselves, documentation should be easily accessible for the whole community and easy to improve from developers and users alike.


\section{Conclusion}

Computer modeling will remain an essential component of advanced accelerator research for design, operation and interpretation, even more so as the accelerators mature to increasingly complex production tools with stringent operational constraints.
Important grand challenges of the coming years include the accurate, end-to-end description of multi-stage collider designs with AAC components, the development of hybrid accelerators that arises with upgrades of beamlines with, e.g., plasma focusing elements or boosters. 

Fast turn-around and scalable modeling capabilities are indispensable, yet still often far from providing real-time feedback that could benefit interactive design and guide experiments, especially with sufficient fidelity.
Thus, training of AI/ML surrogate models adds to the need to accelerate modeling capabilities with modern computing hardware and algorithms.
HPC computing trends based on new hardware, most notably GPUs, can contribute to improving the time-to-solution and accuracy of simulations.
With a higher level of parallelism and software complexity, deeper software stacks are needed than in traditional modeling approaches and community libraries emerge around themes such as performance-portability layers, accelerated numerics libraries, multi-node data management, dynamic load balancing, scalable I/O, parallel and in situ data analysis as well as visualization, to name a few.

Ultimately, the realization of such ambitious goals as the realization of start-to-end virtual twins of particle accelerators that can provide feedback in real time calls for (a) a more effective coordination of the codes development and maintenance across the community - which should be encouraged among laboratories, academia, and industrial partners - together with (b) the development of ecosystems of modeling tools for multiphysics particle accelerator modeling and design, standards and end-to-end workflows. To maximize efficiency and benefits to the community, these should be developed using modern software best practices and adopt open science principles to the largest extent possible, with renewed attention on usability, reliability and sustainability. 



\section*{Acknowledgements}

We acknowledge the support of the Director, Office of Science, Office of High Energy Physics, of the
U.S. Department of Energy under Contract No. DEAC02-05CH11231, including support to the Consortium for Advanced Modeling of Particle Accelerators (CAMPA) project, and of the Exascale
Computing Project (No. 17-SC-20-SC), a collaborative effort of the U.S. Department of Energy's Office of
Science and National Nuclear Security Administration. 
We also acknowledge support by the U.S. Department of Energy, Office of Science, Office of High Energy Physics under Award Number DE-SC0018719.
We acknowledge funding by the Gordon and Betty Moore Foundation under grant GMBF4744 (ACHIP) and the German Ministry of Education and Research (Grant No. 05K19RDE). 
We acknowledge support by the U.S. Department of Energy, Office of Science under contracts DE-SC0018656 with Northern Illinois University and DE-AC02-06CH11357 with Argonne National Laboratory.
This research used resources of the National Energy Research Scientific Computing Center (NERSC), a U.S. Department of Energy Office of Science User Facility located at Lawrence Berkeley National Laboratory, operated under Contract No. DE-AC02-05CH11231.
We acknowledge the support by the US National Science Foundation under award number PHY-1505858, as well as the Bose Foundation.

\bibliographystyle{unsrtnat}
\bibliography{references,references_mendeley}

\begin{thebibliography}{149}
\providecommand{\natexlab}[1]{#1}
\providecommand{\url}[1]{\texttt{#1}}
\expandafter\ifx\csname urlstyle\endcsname\relax
  \providecommand{\doi}[1]{doi: #1}\else
  \providecommand{\doi}{doi: \begingroup \urlstyle{rm}\Url}\fi

\bibitem[{The CST Team}()]{CST}
{The CST Team}.
\newblock {CST}.
\newblock URL
  \url{https://www.3ds.com/products-services/simulia/products/cst-studio-suite/}.

\bibitem[{The VSIM Team}()]{VSIM}
{The VSIM Team}.
\newblock {VSIM}.
\newblock URL \url{https://txcorp.com/vsim/}.

\bibitem[Sagan et~al.(2021)Sagan, Cook, Hao, Hoffstaetter, Huebl, Huang,
  Langston, Mayes, Mitchell, Ng, Qiang, Ryne, E.~Stern, Vay, Winklehner, and
  Zhang]{SaganICFA2021}
D.~Sagan, N.~M. Cook, Y.~Hao, G.~Hoffstaetter, A.~Huebl, C.-K. Huang, M.~H.
  Langston, C.~E. Mayes, C.~E. Mitchell, C.-K. Ng, J.~Qiang, R.~D. Ryne,
  E.~E.~Stern, J.-L. Vay, D.~Winklehner, and H.~Zhang.
\newblock {Simulations of Future Particle Accelerators: Issues and
  Mitigations}.
\newblock \emph{This Issue}, 2021.

\bibitem[Chen et~al.(1985)Chen, Dawson, Huff, and Katsouleas]{ChenPRL1985}
Pisin Chen, J.~M. Dawson, Robert~W. Huff, and T.~Katsouleas.
\newblock {Acceleration of Electrons by the Interaction of a Bunched Electron
  Beam with a Plasma}.
\newblock \emph{Physical Review Letters}, 54\penalty0 (7):\penalty0 693, 2
  1985.
\newblock \doi{10.1103/PhysRevLett.54.693}.
\newblock URL
  \url{https://journals.aps.org/prl/abstract/10.1103/PhysRevLett.54.693}.

\bibitem[Esarey et~al.(2009)Esarey, Schroeder, and Leemans]{Esareyrmp09}
E~Esarey, C~B Schroeder, and W~P Leemans.
\newblock {Physics Of Laser-Driven Plasma-Based Electron Accelerators}.
\newblock \emph{Rev. Mod. Phys.}, 81\penalty0 (3):\penalty0 1229--1285, 2009.
\newblock ISSN 0034-6861.
\newblock \doi{10.1103/Revmodphys.81.1229}.

\bibitem[O'Shea et~al.(2020)O'Shea, Andonian, Baturin, Clarke, Hoang, Hogan,
  Naranjo, Williams, Yakimenko, and Rosenzweig]{OShea_2020}
Brendan~D. O'Shea, Gerard Andonian, S.~S. Baturin, Christine~I. Clarke, P.~D.
  Hoang, Mark~J. Hogan, Brian Naranjo, Oliver~B. Williams, Vitaly Yakimenko,
  and James~B. Rosenzweig.
\newblock Suppression of deflecting forces in planar-symmetric dielectric
  wakefield accelerating structures with elliptical bunches.
\newblock \emph{Phys. Rev. Lett.}, 124:\penalty0 104801, Mar 2020.
\newblock \doi{10.1103/PhysRevLett.124.104801}.
\newblock URL \url{https://link.aps.org/doi/10.1103/PhysRevLett.124.104801}.

\bibitem[Gao et~al.(2018)Gao, Ha, Jing, Antipov, Power, Conde, Gai, Chen, Shi,
  Wisniewski, Doran, Liu, Whiteford, Zholents, Piot, and Baturin]{Gao_2018}
Q.~Gao, G.~Ha, C.~Jing, S.~P. Antipov, J.~G. Power, M.~Conde, W.~Gai, H.~Chen,
  J.~Shi, E.~E. Wisniewski, D.~S. Doran, W.~Liu, C.~E. Whiteford, A.~Zholents,
  P.~Piot, and S.~S. Baturin.
\newblock Observation of high transformer ratio of shaped bunch generated by an
  emittance-exchange beam line.
\newblock \emph{Phys. Rev. Lett.}, 120:\penalty0 114801, Mar 2018.
\newblock \doi{10.1103/PhysRevLett.120.114801}.
\newblock URL \url{https://link.aps.org/doi/10.1103/PhysRevLett.120.114801}.

\bibitem[Andonian et~al.(2014)Andonian, Williams, Barber, Bruhwiler, Favier,
  Fedurin, Fitzmorris, Fukasawa, Hoang, Kusche, Naranjo, O'Shea, Stoltz,
  Swinson, Valloni, and Rosenzweig]{Andonian_2014}
G.~Andonian, O.~Williams, S.~Barber, D.~Bruhwiler, P.~Favier, M.~Fedurin,
  K.~Fitzmorris, A.~Fukasawa, P.~Hoang, K.~Kusche, B.~Naranjo, B.~O'Shea,
  P.~Stoltz, C.~Swinson, A.~Valloni, and J.~B. Rosenzweig.
\newblock Planar-dielectric-wakefield accelerator structure using
  bragg-reflector boundaries.
\newblock \emph{Phys. Rev. Lett.}, 113:\penalty0 264801, Dec 2014.
\newblock \doi{10.1103/PhysRevLett.113.264801}.
\newblock URL \url{https://link.aps.org/doi/10.1103/PhysRevLett.113.264801}.

\bibitem[Hoang et~al.(2018)Hoang, Andonian, Gadjev, Naranjo, Sakai, Sudar,
  Williams, Fedurin, Kusche, Swinson, Zhang, and Rosenzweig]{Hoang_2018}
P.~D. Hoang, G.~Andonian, I.~Gadjev, B.~Naranjo, Y.~Sakai, N.~Sudar,
  O.~Williams, M.~Fedurin, K.~Kusche, C.~Swinson, P.~Zhang, and J.~B.
  Rosenzweig.
\newblock Experimental characterization of electron-beam-driven wakefield modes
  in a dielectric-woodpile cartesian symmetric structure.
\newblock \emph{Phys. Rev. Lett.}, 120:\penalty0 164801, Apr 2018.
\newblock \doi{10.1103/PhysRevLett.120.164801}.
\newblock URL \url{https://link.aps.org/doi/10.1103/PhysRevLett.120.164801}.

\bibitem[Lemery et~al.(2018)Lemery, Floettmann, Piot, K\"artner, and
  A\ss{}mann]{Lemery_2018}
F.~Lemery, K.~Floettmann, P.~Piot, F.~X. K\"artner, and R.~A\ss{}mann.
\newblock Synchronous acceleration with tapered dielectric-lined waveguides.
\newblock \emph{Phys. Rev. Accel. Beams}, 21:\penalty0 051302, May 2018.
\newblock \doi{10.1103/PhysRevAccelBeams.21.051302}.
\newblock URL
  \url{https://link.aps.org/doi/10.1103/PhysRevAccelBeams.21.051302}.

\bibitem[Lu et~al.(2019)Lu, Shapiro, Mastovsky, Temkin, Conde, Power, Shao,
  Wisniewski, and Jing]{Lu_2019}
Xueying Lu, Michael~A. Shapiro, Ivan Mastovsky, Richard~J. Temkin, Manoel
  Conde, John~G. Power, Jiahang Shao, Eric~E. Wisniewski, and Chunguang Jing.
\newblock Generation of high-power, reversed-cherenkov wakefield radiation in a
  metamaterial structure.
\newblock \emph{Phys. Rev. Lett.}, 122:\penalty0 014801, Jan 2019.
\newblock \doi{10.1103/PhysRevLett.122.014801}.
\newblock URL \url{https://link.aps.org/doi/10.1103/PhysRevLett.122.014801}.

\bibitem[England et~al.(2014)England, Noble, et~al.]{england:rmp:2014}
R.~J. England, R.~J. Noble, et~al.
\newblock {Dielectric laser accelerators}.
\newblock \emph{Rev. Mod. Phys.}, 86:\penalty0 1337, 2014.

\bibitem[Cesar et~al.(2018)]{cesar:nonlinear:2018}
D.~Cesar et~al.
\newblock {High-field nonlinear optical response and phase control in a
  dielectric laser accelerator}.
\newblock \emph{Communications Physics}, 1:\penalty0 46, 2018.

\bibitem[Niedermayer et~al.(2017)Niedermayer, Egenolf, and
  Boine-Frankenheim]{niedermayer:2017}
U.~Niedermayer, T.~Egenolf, and O.~Boine-Frankenheim.
\newblock {Beam dynamics analysis of dielectric laser acceleration using a fast
  6D tracking scheme}.
\newblock \emph{Phys. Rev. Accel. Beams}, 20\penalty0 (11):\penalty0 111302,
  2017.

\bibitem[Hughes et~al.(2017)]{hughes:avm:2017}
T.~Hughes et~al.
\newblock {Method for computationally efficient design of dielectric laser
  accelerator structures}.
\newblock \emph{Opt. Exp.}, 25\penalty0 (13):\penalty0 15414, 2017.

\bibitem[Birdsall and Langdon(1991)]{Birdsalllangdon}
C~K Birdsall and A~B Langdon.
\newblock \emph{{Plasma Physics Via Computer Simulation}}.
\newblock Adam-Hilger, 1991.
\newblock ISBN 0 07 005371 5.

\bibitem[Lifschitz et~al.(2009)Lifschitz, Davoine, Lefebvre, Faure, Rechatin,
  and Malka]{LifschitzJCP2009}
A~F Lifschitz, X~Davoine, E~Lefebvre, J~Faure, C~Rechatin, and V~Malka.
\newblock {Particle-in-Cell modelling of laser plasma interaction using Fourier
  decomposition}.
\newblock \emph{Journal of Computational Physics}, 228\penalty0 (5):\penalty0
  1803--1814, 2009.
\newblock ISSN 0021-9991.
\newblock \doi{http://dx.doi.org/10.1016/j.jcp.2008.11.017}.
\newblock URL
  \url{http://www.sciencedirect.com/science/article/pii/S0021999108005950}.

\bibitem[Sprangle et~al.(1990)Sprangle, Esarey, and Ting]{Sprangleprl90}
P~Sprangle, E~Esarey, and A~Ting.
\newblock {Nonlinear-Theory Of Intense Laser-Plasma Interactions}.
\newblock \emph{Physical Review Letters}, 64\penalty0 (17):\penalty0
  2011--2014, 4 1990.
\newblock ISSN 0031-9007.

\bibitem[Antonsen and Mora(1992)]{Antonsenpop1997}
T~M Antonsen and P~Mora.
\newblock {Self-Focusing And Raman-Scattering Of Laser-Pulses In Tenuous
  Plasmas}.
\newblock \emph{Physics of Plasmas}, 69:\penalty0 2204--2207, 1992.

\bibitem[Gordon et~al.(2000)Gordon, Mori, and Antonsen]{GordonIEEE2000}
Daniel~F Gordon, W~B Mori, and Thomas~M Antonsen.
\newblock {A Ponderomotive Guiding Center Particle-in-Cell Code for Efficient
  Modeling of Laser–Plasma Interactions}.
\newblock \emph{IEEE TRANSACTIONS ON PLASMA SCIENCE}, 28\penalty0 (4), 2000.

\bibitem[Benedetti et~al.(2010)Benedetti, Schroeder, Esarey, Geddes, and
  Leemans]{Benedettiaac2010}
C~Benedetti, C~B Schroeder, E~Esarey, C~G~R Geddes, and W~P Leemans.
\newblock {Efficient Modeling Of Laser-Plasma Accelerators With Inf{\&}Rno}.
\newblock \emph{Aip Conference Proceedings}, 1299:\penalty0 250--255, 2010.
\newblock \doi{10.1063/1.3520323}.

\bibitem[Cowan et~al.(2011)Cowan, Bruhwiler, Cormier-Michel, Esarey, Geddes,
  Messmer, and Paul]{Cowanjcp11}
Benjamin~M Cowan, David~L Bruhwiler, Estelle Cormier-Michel, Eric Esarey,
  Cameron G~R Geddes, Peter Messmer, and Kevin~M Paul.
\newblock {Characteristics Of An Envelope Model For Laser-Plasma Accelerator
  Simulation}.
\newblock \emph{Journal of Computational Physics}, 230\penalty0 (1):\penalty0
  61--86, 2011.
\newblock ISSN 0021-9991.
\newblock \doi{Doi: 10.1016/J.Jcp.2010.09.009}.

\bibitem[Terzani et~al.(2021)Terzani, Benedetti, Schroeder, and
  Esarey]{TerzaniPoP2021}
D.~Terzani, C.~Benedetti, C.~B. Schroeder, and E.~Esarey.
\newblock {Accuracy of the time-averaged ponderomotive approximation for
  laser-plasma accelerator modeling}.
\newblock \emph{Physics of Plasmas}, 28\penalty0 (6):\penalty0 063105, 6 2021.
\newblock ISSN 1070-664X.
\newblock \doi{10.1063/5.0050580}.
\newblock URL \url{https://aip.scitation.org/doi/abs/10.1063/5.0050580}.

\bibitem[Vay(2007)]{Vayprl07}
J.-L. Vay.
\newblock {Noninvariance Of Space- And Time-Scale Ranges Under A Lorentz
  Transformation And The Implications For The Study Of Relativistic
  Interactions}.
\newblock \emph{Physical Review Letters}, 98\penalty0 (13):\penalty0 1--4,
  2007.
\newblock ISSN 0031-9007.

\bibitem[Vay and Lehe(2017)]{VayRAST2017}
Jean-Luc Vay and Rémi Lehe.
\newblock {Simulations for Plasma and Laser Acceleration}.
\newblock \emph{http://dx.doi.org/10.1142/S1793626816300085}, 9:\penalty0
  165--186, 2 2017.
\newblock \doi{10.1142/S1793626816300085}.

\bibitem[Oskooi et~al.(2010)Oskooi, Roundy, Ibanescu, Bermel, Joannopoulos, and
  Johnson]{meep}
Ardavan~F. Oskooi, David Roundy, Mihai Ibanescu, Peter Bermel, J.D.
  Joannopoulos, and Steven~G. Johnson.
\newblock Meep: A flexible free-software package for electromagnetic
  simulations by the fdtd method.
\newblock \emph{Computer Physics Communications}, 181\penalty0 (3):\penalty0
  687--702, 2010.
\newblock ISSN 0010-4655.
\newblock \doi{https://doi.org/10.1016/j.cpc.2009.11.008}.
\newblock URL
  \url{https://www.sciencedirect.com/science/article/pii/S001046550900383X}.

\bibitem[Vay et~al.(2012{\natexlab{a}})Vay, Grote, Cohen, and
  Friedman]{VayCSD12}
J.-L. Vay, D~P Grote, R~H Cohen, and A~Friedman.
\newblock {Novel methods in the particle-in-cell accelerator code-framework
  warp}.
\newblock \emph{Computational Science and Discovery}, 5\penalty0 (1):\penalty0
  014019 (20 pp.), 2012{\natexlab{a}}.
\newblock ISSN 1749-4680.

\bibitem[Borland(2000)]{elegant}
M.~Borland.
\newblock {elegant: A Flexible SDDS-Compliant Code for Accelerator Simulation}.
\newblock In \emph{{6th International Computational Accelerator Physics
  Conference (ICAP 2000)}}, 2000.
\newblock \doi{10.2172/761286}.

\bibitem[Floettmann et~al.(2003)Floettmann, Lidia, and Piot]{astra}
K.~Floettmann, S.~M. Lidia, and P.~Piot.
\newblock {Recent Improvements to the ASTRA Particle Tracking Code}.
\newblock \emph{Conf. Proc. C}, 030512:\penalty0 3500, 2003.

\bibitem[Bane and Emma(2005)]{litrack}
K.~L.~F. Bane and P.~Emma.
\newblock {LiTrack: A Fast longitudinal phase space tracking code with
  graphical user interface}.
\newblock \emph{Conf. Proc. C}, 0505161:\penalty0 4266, 2005.

\bibitem[Tan et~al.(2021)Tan, Piot, and Zholents]{PhysRevAccelBeams.24.051303}
Wei~Hou Tan, Philippe Piot, and Alexander Zholents.
\newblock Formation of temporally shaped electron bunches for beam-driven
  collinear wakefield accelerators.
\newblock \emph{Phys. Rev. Accel. Beams}, 24:\penalty0 051303, May 2021.
\newblock \doi{10.1103/PhysRevAccelBeams.24.051303}.
\newblock URL
  \url{https://link.aps.org/doi/10.1103/PhysRevAccelBeams.24.051303}.

\bibitem[Mihalcea et~al.(2012)Mihalcea, Piot, and
  Stoltz]{PhysRevSTAB.15.081304}
D.~Mihalcea, P.~Piot, and P.~Stoltz.
\newblock Three-dimensional analysis of wakefields generated by flat electron
  beams in planar dielectric-loaded structures.
\newblock \emph{Phys. Rev. ST Accel. Beams}, 15:\penalty0 081304, Aug 2012.
\newblock \doi{10.1103/PhysRevSTAB.15.081304}.
\newblock URL \url{https://link.aps.org/doi/10.1103/PhysRevSTAB.15.081304}.

\bibitem[Pacey et~al.(2018)Pacey, Saveliev, Xia, and Smith]{Pacey:2017ynk}
Thomas~H. Pacey, Yuri Saveliev, Guoxing Xia, and Jonathan Smith.
\newblock {Simulation studies for dielectric wakefield programme at CLARA
  facility}.
\newblock \emph{Nucl. Instrum. Meth. A}, 909:\penalty0 261--265, 2018.
\newblock \doi{10.1016/j.nima.2017.12.038}.

\bibitem[Zagorodnov(2014)]{Zagorodnov:206299}
Igor Zagorodnov.
\newblock {W}akefield {C}ode {ECHO} 2(3){D}.
\newblock ICFA mini-Workshop on "Electromagnetic wake fields and impedances in
  particle accelerators", Erice, Sicily (Italy), 24 Apr 2014 - 28 Apr 2014, Apr
  2014.
\newblock URL \url{https://bib-pubdb1.desy.de/record/206299}.

\bibitem[ace()]{ace3p}
{ACE3P}.
\newblock URL
  \url{https://portal.slac.stanford.edu/sites/ard_public/acd/Pages/Default.aspx}.

\bibitem[Vincenti and Vay(2018)]{VincentiCPC2018}
H.~Vincenti and J.-L. Vay.
\newblock {Ultrahigh-order Maxwell solver with extreme scalability for
  electromagnetic PIC simulations of plasmas}.
\newblock \emph{Computer Physics Communications}, 2018.
\newblock ISSN 00104655.
\newblock \doi{10.1016/j.cpc.2018.03.018}.

\bibitem[O'Shea et~al.(2019)O'Shea, Andonian, Barber, Clarke, Hoang, Hogan,
  Naranjo, Williams, Yakimenko, and Rosenzweig]{PhysRevLett.123.134801}
B.~D. O'Shea, G.~Andonian, S.~K. Barber, C.~I. Clarke, P.~D. Hoang, M.~J.
  Hogan, B.~Naranjo, O.~B. Williams, V.~Yakimenko, and J.~B. Rosenzweig.
\newblock Conductivity induced by high-field terahertz waves in dielectric
  material.
\newblock \emph{Phys. Rev. Lett.}, 123:\penalty0 134801, Sep 2019.
\newblock \doi{10.1103/PhysRevLett.123.134801}.
\newblock URL \url{https://link.aps.org/doi/10.1103/PhysRevLett.123.134801}.

\bibitem[{ALEGRO}(2019)]{ALEGRO}
{ALEGRO}.
\newblock {Towards an Advanced Linear International Collider}.
\newblock 1 2019.
\newblock URL \url{http://arxiv.org/abs/1901.10370}.

\bibitem[Shin and Fan(2013)]{shin:2013}
W.~Shin and S.~Fan.
\newblock {Accelerated solution of the frequency-domain Maxwell's equations by
  enginnering the eigenvalue distribution of the operator}.
\newblock \emph{Opt. Exp.}, 21\penalty0 (19):\penalty0 22578, 2013.

\bibitem[Egenolf et~al.(2017)Egenolf, Boine-Frankenheim, and
  Niedermayer]{egenolf:2017}
T.~Egenolf, O.~Boine-Frankenheim, and U.~Niedermayer.
\newblock {Simulation of DLA Grating Structures in the Frequency Domain}.
\newblock \emph{Journal of Physics: Conference Series.}, 874\penalty0
  (1):\penalty0 012040, 2017.

\bibitem[Niedermayer et~al.(2018)Niedermayer, Egenolf, Boine-Frankenheim, and
  Hommelhoff]{niedermayerPRL2018}
Uwe Niedermayer, Thilo Egenolf, Oliver Boine-Frankenheim, and Peter Hommelhoff.
\newblock {Alternating-Phase Focusing for Dielectric-Laser Acceleration}.
\newblock \emph{Physical Review Letters}, 121\penalty0 (21):\penalty0 214801,
  11 2018.
\newblock \doi{10.1103/PhysRevLett.121.214801}.
\newblock URL
  \url{https://journals.aps.org/prl/abstract/10.1103/PhysRevLett.121.214801}.

\bibitem[Niedermayer et~al.(2020)Niedermayer, Egenolf, and
  Boine-Frankenheim]{niedermayerPRL2020}
Uwe Niedermayer, Thilo Egenolf, and Oliver Boine-Frankenheim.
\newblock {Three Dimensional Alternating-Phase Focusing for Dielectric-Laser
  Electron Accelerators}.
\newblock \emph{Physical Review Letters}, 125\penalty0 (16):\penalty0 164801,
  10 2020.
\newblock \doi{10.1103/PhysRevLett.125.164801}.
\newblock URL
  \url{https://journals.aps.org/prl/abstract/10.1103/PhysRevLett.125.164801}.

\bibitem[Egenolf et~al.(2020)Egenolf, Niedermayer, and
  Boine-Frankenheim]{egenolfPRAB2020}
Thilo Egenolf, Uwe Niedermayer, and Oliver Boine-Frankenheim.
\newblock {Tracking with wakefields in dielectric laser acceleration grating
  structures}.
\newblock \emph{Physical Review Accelerators and Beams}, 23\penalty0
  (5):\penalty0 054402, 5 2020.
\newblock \doi{10.1103/PhysRevAccelBeams.23.054402}.
\newblock URL
  \url{https://journals.aps.org/prab/abstract/10.1103/PhysRevAccelBeams.23.054402}.

\bibitem[Geddes et~al.(2004)Geddes, Toth, van Tilborg, Esarey, Schroeder,
  Bruhwiler, Nieter, Cary, and Leemans]{Geddes_2004}
C.~G.~R. Geddes, Cs. Toth, J.~van Tilborg, E.~Esarey, C.~B. Schroeder,
  D.~Bruhwiler, C.~Nieter, J.~Cary, and W.~P. Leemans.
\newblock High-quality electron beams from a laser wakefield accelerator using
  plasma-channel guiding.
\newblock \emph{Nature}, 431\penalty0 (7008):\penalty0 538--541, 09 2004.
\newblock URL \url{http://dx.doi.org/10.1038/nature02900}.

\bibitem[Leemans et~al.(2006)Leemans, Nagler, Gonsalves, Toth, Nakamura,
  Geddes, Esarey, Schroeder, and Hooker]{Leemans_2006}
W.~P. Leemans, B.~Nagler, A.~J. Gonsalves, Cs. Toth, K.~Nakamura, C.~G.~R.
  Geddes, E.~Esarey, C.~B. Schroeder, and S.~M. Hooker.
\newblock Gev electron beams from a centimetre-scale accelerator.
\newblock \emph{Nat Phys}, 2\penalty0 (10):\penalty0 696--699, 10 2006.
\newblock URL \url{http://dx.doi.org/10.1038/nphys418}.

\bibitem[Ibbotson et~al.(2010)Ibbotson, Bourgeois, Rowlands-Rees, Caballero,
  Bajlekov, Walker, Kneip, Mangles, Nagel, Palmer, Delerue, Doucas, Urner,
  Chekhlov, Clarke, Divall, Ertel, Foster, Hawkes, Hooker, Parry, Rajeev,
  Streeter, and Hooker]{Ibbotson_2010}
T.~P.~A. Ibbotson, N.~Bourgeois, T.~P. Rowlands-Rees, L.~S. Caballero, S.~I.
  Bajlekov, P.~A. Walker, S.~Kneip, S.~P.~D. Mangles, S.~R. Nagel, C.~A.~J.
  Palmer, N.~Delerue, G.~Doucas, D.~Urner, O.~Chekhlov, R.~J. Clarke,
  E.~Divall, K.~Ertel, P.~S. Foster, S.~J. Hawkes, C.~J. Hooker, B.~Parry,
  P.~P. Rajeev, M.~J.~V. Streeter, and S.~M. Hooker.
\newblock Laser-wakefield acceleration of electron beams in a low density
  plasma channel.
\newblock \emph{Phys. Rev. ST Accel. Beams}, 13:\penalty0 031301, Mar 2010.
\newblock \doi{10.1103/PhysRevSTAB.13.031301}.
\newblock URL \url{http://link.aps.org/doi/10.1103/PhysRevSTAB.13.031301}.

\bibitem[Suk et~al.(2001)Suk, Barov, Rosenzweig, and Esarey]{Suk_2001}
H.~Suk, N.~Barov, J.~B. Rosenzweig, and E.~Esarey.
\newblock Plasma electron trapping and acceleration in a plasma wake field
  using a density transition.
\newblock \emph{Phys. Rev. Lett.}, 86:\penalty0 1011--1014, Feb 2001.
\newblock \doi{10.1103/PhysRevLett.86.1011}.
\newblock URL \url{https://link.aps.org/doi/10.1103/PhysRevLett.86.1011}.

\bibitem[Sprangle et~al.(2000)Sprangle, Hafizi, Pe\~nano, Hubbard, Ting,
  Zigler, and Antonsen]{Sprangle_2000}
P.~Sprangle, B.~Hafizi, J.~R. Pe\~nano, R.~F. Hubbard, A.~Ting, A.~Zigler, and
  T.~M. Antonsen.
\newblock Stable laser-pulse propagation in plasma channels for gev electron
  acceleration.
\newblock \emph{Phys. Rev. Lett.}, 85:\penalty0 5110--5113, Dec 2000.
\newblock \doi{10.1103/PhysRevLett.85.5110}.
\newblock URL \url{https://link.aps.org/doi/10.1103/PhysRevLett.85.5110}.

\bibitem[Blue et~al.(2003)Blue, Clayton, O'Connell, Decker, Hogan, Huang,
  Iverson, Joshi, Katsouleas, Lu, Marsh, Mori, Muggli, Siemann, and
  Walz]{Blue_2003}
B.~E. Blue, C.~E. Clayton, C.~L. O'Connell, F.-J. Decker, M.~J. Hogan,
  C.~Huang, R.~Iverson, C.~Joshi, T.~C. Katsouleas, W.~Lu, K.~A. Marsh, W.~B.
  Mori, P.~Muggli, R.~Siemann, and D.~Walz.
\newblock Plasma-wakefield acceleration of an intense positron beam.
\newblock \emph{Phys. Rev. Lett.}, 90:\penalty0 214801, May 2003.
\newblock \doi{10.1103/PhysRevLett.90.214801}.
\newblock URL \url{https://link.aps.org/doi/10.1103/PhysRevLett.90.214801}.

\bibitem[Blumenfeld et~al.(2007)Blumenfeld, Clayton, Decker, Hogan, Huang,
  Ischebeck, Iverson, Joshi, Katsouleas, Kirby, Lu, Marsh, Mori, Muggli, Oz,
  Siemann, Walz, and Zhou]{Blumenfeld_2007}
Ian Blumenfeld, Christopher~E. Clayton, Franz-Josef Decker, Mark~J. Hogan,
  Chengkun Huang, Rasmus Ischebeck, Richard Iverson, Chandrashekhar Joshi,
  Thomas Katsouleas, Neil Kirby, Wei Lu, Kenneth~A. Marsh, Warren~B. Mori,
  Patric Muggli, Erdem Oz, Robert~H. Siemann, Dieter Walz, and Miaomiao Zhou.
\newblock Energy doubling of 42 gev electrons in a metre-scale plasma wakefield
  accelerator.
\newblock \emph{Nature}, 445:\penalty0 741 EP --, 02 2007.
\newblock URL \url{http://dx.doi.org/10.1038/nature05538}.

\bibitem[Litos et~al.(2014)Litos, Adli, An, Clarke, Clayton, Corde, Delahaye,
  England, Fisher, Frederico, Gessner, Green, Hogan, Joshi, Lu, Marsh, Mori,
  Muggli, Vafaei-Najafabadi, Walz, White, Wu, Yakimenko, and Yocky]{Litos_2014}
M.~Litos, E.~Adli, W.~An, C.~I. Clarke, C.~E. Clayton, S.~Corde, J.~P.
  Delahaye, R.~J. England, A.~S. Fisher, J.~Frederico, S.~Gessner, S.~Z. Green,
  M.~J. Hogan, C.~Joshi, W.~Lu, K.~A. Marsh, W.~B. Mori, P.~Muggli,
  N.~Vafaei-Najafabadi, D.~Walz, G.~White, Z.~Wu, V.~Yakimenko, and G.~Yocky.
\newblock High-efficiency acceleration of an electron beam in a plasma
  wakefield accelerator.
\newblock \emph{Nature}, 515:\penalty0 92 EP --, 11 2014.
\newblock URL \url{http://dx.doi.org/10.1038/nature13882}.

\bibitem[Hogan et~al.(2010)Hogan, Raubenheimer, Seryi, Muggli, Katsouleas,
  Huang, Lu, An, Marsh, Mori, Clayton, and Joshi]{Hogan_2010}
M~J Hogan, T~O Raubenheimer, A~Seryi, P~Muggli, T~Katsouleas, C~Huang, W~Lu,
  W~An, K~A Marsh, W~B Mori, C~E Clayton, and C~Joshi.
\newblock Plasma wakefield acceleration experiments at facet.
\newblock \emph{New Journal of Physics}, 12\penalty0 (5):\penalty0 055030,
  2010.
\newblock URL \url{http://stacks.iop.org/1367-2630/12/i=5/a=055030}.

\bibitem[Green et~al.(2014)Green, Adli, Clarke, Corde, Edstrom, Fisher,
  Frederico, Frisch, Gessner, Gilevich, Hering, Hogan, Jobe, Litos, May, Walz,
  Yakimenko, Clayton, Joshi, Marsh, Vafaei-Najafabadi, and Muggli]{Green_2014}
S~Z Green, E~Adli, C~I Clarke, S~Corde, S~A Edstrom, A~S Fisher, J~Frederico,
  J~C Frisch, S~Gessner, S~Gilevich, P~Hering, M~J Hogan, R~K Jobe, M~Litos,
  J~E May, D~R Walz, V~Yakimenko, C~E Clayton, C~Joshi, K~A Marsh,
  N~Vafaei-Najafabadi, and P~Muggli.
\newblock Laser ionized preformed plasma at facet.
\newblock \emph{Plasma Physics and Controlled Fusion}, 56\penalty0
  (8):\penalty0 084011, 2014.
\newblock URL \url{http://stacks.iop.org/0741-3335/56/i=8/a=084011}.

\bibitem[Gessner et~al.(2016)Gessner, Adli, Allen, An, Clarke, Clayton, Corde,
  Delahaye, Frederico, Green, Hast, Hogan, Joshi, Lindstr{\o}m, Lipkowitz,
  Litos, Lu, Marsh, Mori, O'Shea, Vafaei-Najafabadi, Walz, Yakimenko, and
  Yocky]{Gessner_2016}
Spencer Gessner, Erik Adli, James~M. Allen, Weiming An, Christine~I. Clarke,
  Chris~E. Clayton, Sebastien Corde, J.~P. Delahaye, Joel Frederico, Selina~Z.
  Green, Carsten Hast, Mark~J. Hogan, Chan Joshi, Carl~A. Lindstr{\o}m, Nate
  Lipkowitz, Michael Litos, Wei Lu, Kenneth~A. Marsh, Warren~B. Mori, Brendan
  O'Shea, Navid Vafaei-Najafabadi, Dieter Walz, Vitaly Yakimenko, and Gerald
  Yocky.
\newblock Demonstration of a positron beam-driven hollow channel plasma
  wakefield accelerator.
\newblock \emph{Nature Communications}, 7:\penalty0 11785 EP --, 06 2016.
\newblock URL \url{http://dx.doi.org/10.1038/ncomms11785}.

\bibitem[Lee et~al.(2001)Lee, Katsouleas, Hemker, Dodd, and Mori]{Lee_2001}
S.~Lee, T.~Katsouleas, R.~G. Hemker, E.~S. Dodd, and W.~B. Mori.
\newblock Plasma-wakefield acceleration of a positron beam.
\newblock \emph{Phys. Rev. E}, 64:\penalty0 045501, Sep 2001.
\newblock \doi{10.1103/PhysRevE.64.045501}.
\newblock URL \url{https://link.aps.org/doi/10.1103/PhysRevE.64.045501}.

\bibitem[Schroeder et~al.(2013)Schroeder, Esarey, Benedetti, and
  Leemans]{Schroeder_2013}
C.~B. Schroeder, E.~Esarey, C.~Benedetti, and W.~P. Leemans.
\newblock Control of focusing forces and emittances in plasma-based
  accelerators using near-hollow plasma channels.
\newblock \emph{Physics of Plasmas}, 20\penalty0 (8):\penalty0 080701, 02 2013.
\newblock \doi{10.1063/1.4817799}.
\newblock URL \url{https://doi.org/10.1063/1.4817799}.

\bibitem[Diederichs et~al.(2019)Diederichs, Mehrling, Benedetti, Schroeder,
  Knetsch, Esarey, and Osterhoff]{Diederichs_2019}
S.~Diederichs, T.~J. Mehrling, C.~Benedetti, C.~B. Schroeder, A.~Knetsch,
  E.~Esarey, and J.~Osterhoff.
\newblock Positron transport and acceleration in beam-driven plasma wakefield
  accelerators using plasma columns.
\newblock \emph{Phys. Rev. Accel. Beams}, 22:\penalty0 081301, Aug 2019.
\newblock \doi{10.1103/PhysRevAccelBeams.22.081301}.
\newblock URL
  \url{https://link.aps.org/doi/10.1103/PhysRevAccelBeams.22.081301}.

\bibitem[van Tilborg et~al.(2015)van Tilborg, Steinke, Geddes, Matlis, Shaw,
  Gonsalves, Huijts, Nakamura, Daniels, Schroeder, Benedetti, Esarey, Bulanov,
  Bobrova, Sasorov, and Leemans]{vanTilborg_2015}
J.~van Tilborg, S.~Steinke, C.~G.~R. Geddes, N.~H. Matlis, B.~H. Shaw, A.~J.
  Gonsalves, J.~V. Huijts, K.~Nakamura, J.~Daniels, C.~B. Schroeder,
  C.~Benedetti, E.~Esarey, S.~S. Bulanov, N.~A. Bobrova, P.~V. Sasorov, and
  W.~P. Leemans.
\newblock Active plasma lensing for relativistic laser-plasma-accelerated
  electron beams.
\newblock \emph{Phys. Rev. Lett.}, 115:\penalty0 184802, Oct 2015.
\newblock \doi{10.1103/PhysRevLett.115.184802}.
\newblock URL \url{http://link.aps.org/doi/10.1103/PhysRevLett.115.184802}.

\bibitem[Steinke et~al.(2016)Steinke, van Tilborg, Benedetti, Geddes,
  Schroeder, Daniels, Swanson, Gonsalves, Nakamura, Matlis, Shaw, Esarey, and
  Leemans]{Steinke_2016}
S.~Steinke, J.~van Tilborg, C.~Benedetti, C.~G.~R. Geddes, C.~B. Schroeder,
  J.~Daniels, K.~K. Swanson, A.~J. Gonsalves, K.~Nakamura, N.~H. Matlis, B.~H.
  Shaw, E.~Esarey, and W.~P. Leemans.
\newblock Multistage coupling of independent laser-plasma accelerators.
\newblock \emph{Nature}, 530\penalty0 (7589):\penalty0 190--193, 02 2016.
\newblock URL \url{http://dx.doi.org/10.1038/nature16525}.

\bibitem[Lehe et~al.(2014)Lehe, Thaury, Guillaume, Lifschitz, and
  Malka]{Lehe_2014}
R.~Lehe, C.~Thaury, E.~Guillaume, A.~Lifschitz, and V.~Malka.
\newblock Laser-plasma lens for laser-wakefield accelerators.
\newblock \emph{Phys. Rev. ST Accel. Beams}, 17:\penalty0 121301, Dec 2014.
\newblock \doi{10.1103/PhysRevSTAB.17.121301}.
\newblock URL \url{https://link.aps.org/doi/10.1103/PhysRevSTAB.17.121301}.

\bibitem[Doss et~al.(2019)Doss, Adli, Ariniello, Cary, Corde, Hidding, Hogan,
  Hunt-Stone, Joshi, Marsh, Rosenzweig, Vafaei-Najafabadi, Yakimenko, and
  Litos]{Doss_2019}
C.~E. Doss, E.~Adli, R.~Ariniello, J.~Cary, S.~Corde, B.~Hidding, M.~J. Hogan,
  K.~Hunt-Stone, C.~Joshi, K.~A. Marsh, J.~B. Rosenzweig, N.~Vafaei-Najafabadi,
  V.~Yakimenko, and M.~Litos.
\newblock Laser-ionized, beam-driven, underdense, passive thin plasma lens.
\newblock \emph{Phys. Rev. Accel. Beams}, 22:\penalty0 111001, Nov 2019.
\newblock \doi{10.1103/PhysRevAccelBeams.22.111001}.
\newblock URL
  \url{https://link.aps.org/doi/10.1103/PhysRevAccelBeams.22.111001}.

\bibitem[D'Arcy et~al.(2019)D'Arcy, Wesch, Aschikhin, Bohlen, Behrens, Garland,
  Goldberg, Gonzalez, Knetsch, Libov, de~la Ossa, Meisel, Mehrling, Niknejadi,
  Poder, R\"ockemann, Schaper, Schmidt, Schr\"oder, Palmer, Schwinkendorf,
  Sheeran, Streeter, Tauscher, Wacker, and Osterhoff]{DArcy_2019}
R.~D'Arcy, S.~Wesch, A.~Aschikhin, S.~Bohlen, C.~Behrens, M.~J. Garland,
  L.~Goldberg, P.~Gonzalez, A.~Knetsch, V.~Libov, A.~Martinez de~la Ossa,
  M.~Meisel, T.~J. Mehrling, P.~Niknejadi, K.~Poder, J.-H. R\"ockemann,
  L.~Schaper, B.~Schmidt, S.~Schr\"oder, C.~Palmer, J.-P. Schwinkendorf,
  B.~Sheeran, M.~J.~V. Streeter, G.~Tauscher, V.~Wacker, and J.~Osterhoff.
\newblock Tunable plasma-based energy dechirper.
\newblock \emph{Phys. Rev. Lett.}, 122:\penalty0 034801, Jan 2019.
\newblock \doi{10.1103/PhysRevLett.122.034801}.
\newblock URL \url{https://link.aps.org/doi/10.1103/PhysRevLett.122.034801}.

\bibitem[Scherkl et~al.(2019)Scherkl, Knetsch, Heinemann, Sutherland, Habib,
  Karger, Ullmann, Beaton, Kirwan, Manahan, Xi, Deng, Litos, OShea, Green,
  Clarke, Andonian, Assmann, Jaroszynski, Bruhwiler, Smith, Cary, Hogan,
  Yakimenko, Rosenzweig, and Hidding]{Scherkl_2019}
Paul Scherkl, Alexander Knetsch, Thomas Heinemann, Andrew Sutherland,
  Ahmad~Fahim Habib, Oliver Karger, Daniel Ullmann, Andrew Beaton, Gavin
  Kirwan, Grace Manahan, Yunfeng Xi, Aihua Deng, Michael~Dennis Litos,
  Brendan~D. OShea, Selina~Z. Green, Christine~I. Clarke, Gerard Andonian,
  Ralph Assmann, Dino~A. Jaroszynski, David~L. Bruhwiler, Jonathan Smith,
  John~R. Cary, Mark~J. Hogan, Vitaly Yakimenko, James~B. Rosenzweig, and
  Bernhard Hidding.
\newblock Plasma-photonic spatiotemporal synchronization of relativistic
  electron and laser beams, 2019.

\bibitem[Bagdasarov et~al.(2017)Bagdasarov, Sasorov, Gasilov, Boldarev,
  Olkhovskaya, Benedetti, Bulanov, Gonsalves, Mao, Schroeder, van Tilborg,
  Esarey, Leemans, Levato, Margarone, and Korn]{Bagdasarov_2017}
G.~A. Bagdasarov, P.~V. Sasorov, V.~A. Gasilov, A.~S. Boldarev, O.~G.
  Olkhovskaya, C.~Benedetti, S.~S. Bulanov, A.~Gonsalves, H.~S. Mao, C.~B.
  Schroeder, J.~van Tilborg, E.~Esarey, W.~P. Leemans, T.~Levato, D.~Margarone,
  and G.~Korn.
\newblock Laser beam coupling with capillary discharge plasma for laser
  wakefield acceleration applications.
\newblock \emph{Physics of Plasmas}, 24\penalty0 (8):\penalty0 083109,
  2018/02/20 2017.
\newblock \doi{10.1063/1.4997606}.
\newblock URL \url{https://doi.org/10.1063/1.4997606}.

\bibitem[Gonsalves et~al.(2019)Gonsalves, Nakamura, Daniels, Benedetti,
  Pieronek, de~Raadt, Steinke, Bin, Bulanov, van Tilborg, Geddes, Schroeder,
  T\'oth, Esarey, Swanson, Fan-Chiang, Bagdasarov, Bobrova, Gasilov, Korn,
  Sasorov, and Leemans]{Gonsalves_2019}
A.~J. Gonsalves, K.~Nakamura, J.~Daniels, C.~Benedetti, C.~Pieronek, T.~C.~H.
  de~Raadt, S.~Steinke, J.~H. Bin, S.~S. Bulanov, J.~van Tilborg, C.~G.~R.
  Geddes, C.~B. Schroeder, Cs. T\'oth, E.~Esarey, K.~Swanson, L.~Fan-Chiang,
  G.~Bagdasarov, N.~Bobrova, V.~Gasilov, G.~Korn, P.~Sasorov, and W.~P.
  Leemans.
\newblock Petawatt laser guiding and electron beam acceleration to 8 gev in a
  laser-heated capillary discharge waveguide.
\newblock \emph{Phys. Rev. Lett.}, 122:\penalty0 084801, 2 2019.
\newblock \doi{10.1103/PhysRevLett.122.084801}.
\newblock URL \url{https://link.aps.org/doi/10.1103/PhysRevLett.122.084801}.

\bibitem[Bagdasarov et~al.(2021)Bagdasarov, Bobrova, Olkhovskaya, Gasilov,
  Benedetti, Bulanov, Gonsalves, Pieronek, van Tilborg, Geddes, Schroeder,
  Sasorov, Bulanov, Korn, and Esarey]{Bagdasarov_2021}
G.~A. Bagdasarov, N.~A. Bobrova, O.~G. Olkhovskaya, V.~A. Gasilov,
  C.~Benedetti, S.~S. Bulanov, A.~J. Gonsalves, C.~V. Pieronek, J.~van Tilborg,
  C.~G.~R. Geddes, C.~B. Schroeder, P.~V. Sasorov, S.~V. Bulanov, G.~Korn, and
  E.~Esarey.
\newblock Creation of an axially uniform plasma channel in a laser-assisted
  capillary discharge.
\newblock \emph{Physics of Plasmas}, 28\penalty0 (5):\penalty0 053104, 2021.
\newblock \doi{10.1063/5.0046428}.
\newblock URL \url{https://doi.org/10.1063/5.0046428}.

\bibitem[van Tilborg et~al.(2017)van Tilborg, Barber, Tsai, Swanson, Steinke,
  Geddes, Gonsalves, Schroeder, Esarey, Bulanov, Bobrova, Sasorov, and
  Leemans]{vanTilborg_2017}
J.~van Tilborg, S.~K. Barber, H.-E. Tsai, K.~K. Swanson, S.~Steinke, C.~G.~R.
  Geddes, A.~J. Gonsalves, C.~B. Schroeder, E.~Esarey, S.~S. Bulanov, N.~A.
  Bobrova, P.~V. Sasorov, and W.~P. Leemans.
\newblock Nonuniform discharge currents in active plasma lenses.
\newblock \emph{Phys. Rev. Accel. Beams}, 20:\penalty0 032803, Mar 2017.
\newblock \doi{10.1103/PhysRevAccelBeams.20.032803}.
\newblock URL
  \url{https://link.aps.org/doi/10.1103/PhysRevAccelBeams.20.032803}.

\bibitem[Lindstr\o{}m et~al.(2018)Lindstr\o{}m, Adli, Boyle, Corsini, Dyson,
  Farabolini, Hooker, Meisel, Osterhoff, R\"ockemann, Schaper, and
  Sjobak]{Lindstrom_2018}
C.~A. Lindstr\o{}m, E.~Adli, G.~Boyle, R.~Corsini, A.~E. Dyson, W.~Farabolini,
  S.~M. Hooker, M.~Meisel, J.~Osterhoff, J.-H. R\"ockemann, L.~Schaper, and
  K.~N. Sjobak.
\newblock Emittance preservation in an aberration-free active plasma lens.
\newblock \emph{Phys. Rev. Lett.}, 121:\penalty0 194801, Nov 2018.
\newblock \doi{10.1103/PhysRevLett.121.194801}.
\newblock URL \url{https://link.aps.org/doi/10.1103/PhysRevLett.121.194801}.

\bibitem[Cook et~al.(2020)Cook, Carlsson, Moeller, Nagler, and
  Tzeferacos]{Cook_2020}
N~M Cook, J~Carlsson, P~Moeller, R~Nagler, and P~Tzeferacos.
\newblock Modeling of capillary discharge plasmas for wakefield acceleration
  and beam transport.
\newblock \emph{Journal of Physics: Conference Series}, 1596:\penalty0 012063,
  sep 2020.
\newblock \doi{10.1088/1742-6596/1596/1/012063}.
\newblock URL \url{https://doi.org/10.1088/1742-6596/1596/1/012063}.

\bibitem[Xu et~al.(2016)Xu, Hua, Wu, Zhang, Li, Wan, Pai, Lu, An, Yu, Hogan,
  Joshi, and Mori]{Xu_2016}
X.~L. Xu, J.~F. Hua, Y.~P. Wu, C.~J. Zhang, F.~Li, Y.~Wan, C.-H. Pai, W.~Lu,
  W.~An, P.~Yu, M.~J. Hogan, C.~Joshi, and W.~B. Mori.
\newblock Physics of phase space matching for staging plasma and traditional
  accelerator components using longitudinally tailored plasma profiles.
\newblock \emph{Phys. Rev. Lett.}, 116:\penalty0 124801, Mar 2016.
\newblock \doi{10.1103/PhysRevLett.116.124801}.
\newblock URL \url{https://link.aps.org/doi/10.1103/PhysRevLett.116.124801}.

\bibitem[Deng et~al.(2019)Deng, Karger, Heinemann, Knetsch, Scherkl, Manahan,
  Beaton, Ullmann, Wittig, Habib, Xi, Litos, O'Shea, Gessner, Clarke, Green,
  Lindstr{\o}m, Adli, Zgadzaj, Downer, Andonian, Murokh, Bruhwiler, Cary,
  Hogan, Yakimenko, Rosenzweig, and Hidding]{Deng_2019}
A.~Deng, O.~S. Karger, T.~Heinemann, A.~Knetsch, P.~Scherkl, G.~G. Manahan,
  A.~Beaton, D.~Ullmann, G.~Wittig, A.~F. Habib, Y.~Xi, M.~D. Litos, B.~D.
  O'Shea, S.~Gessner, C.~I. Clarke, S.~Z. Green, C.~A. Lindstr{\o}m, E.~Adli,
  R.~Zgadzaj, M.~C. Downer, G.~Andonian, A.~Murokh, D.~L. Bruhwiler, J.~R.
  Cary, M.~J. Hogan, V.~Yakimenko, J.~B. Rosenzweig, and B.~Hidding.
\newblock Generation and acceleration of electron bunches from a plasma
  photocathode.
\newblock \emph{Nature Physics}, 15\penalty0 (11):\penalty0 1156--1160, 2019.
\newblock \doi{10.1038/s41567-019-0610-9}.
\newblock URL \url{https://doi.org/10.1038/s41567-019-0610-9}.

\bibitem[Xi et~al.(2013)Xi, Hidding, Bruhwiler, Pretzler, and
  Rosenzweig]{Xi_2013}
Y.~Xi, B.~Hidding, D.~Bruhwiler, G.~Pretzler, and J.~B. Rosenzweig.
\newblock Hybrid modeling of relativistic underdense plasma photocathode
  injectors.
\newblock \emph{Phys. Rev. ST Accel. Beams}, 16:\penalty0 031303, Mar 2013.
\newblock \doi{10.1103/PhysRevSTAB.16.031303}.
\newblock URL \url{https://link.aps.org/doi/10.1103/PhysRevSTAB.16.031303}.

\bibitem[Manahan et~al.(2019)Manahan, Habib, Scherkl, Ullmann, Beaton,
  Sutherland, Kirwan, Delinikolas, Heinemann, Altuijri, Knetsch, Karger, Cook,
  Bruhwiler, Sheng, Rosenzweig, and Hidding]{Manahan_2019}
G.~G. Manahan, A.~F. Habib, P.~Scherkl, D.~Ullmann, A.~Beaton, A.~Sutherland,
  G.~Kirwan, P.~Delinikolas, T.~Heinemann, R.~Altuijri, A.~Knetsch, O.~Karger,
  N.~M. Cook, D.~L. Bruhwiler, Z.~M. Sheng, J.~B. Rosenzweig, and B.~Hidding.
\newblock Advanced schemes for underdense plasma photocathode wakefield
  accelerators: pathways towards ultrahigh brightness electron beams.
\newblock \emph{Philosophical Transactions of the Royal Society A:
  Mathematical, Physical and Engineering Sciences}, 377\penalty0
  (2151):\penalty0 20180182, 2020/08/27 2019.
\newblock \doi{10.1098/rsta.2018.0182}.
\newblock URL \url{https://doi.org/10.1098/rsta.2018.0182}.

\bibitem[Gessner and the AWAKE~Collaboration(2020)]{Gessner_2020}
S.~Gessner and the AWAKE~Collaboration.
\newblock Evolution of a plasma column measured through modulation of a
  high-energy proton beam, 2020.

\bibitem[Vay et~al.(2020)Vay, Sagan, Huebl, Th\'evenet, Lehe, Ng, Vincenti,
  Bussmann, Debus, Pausch, and Qiang]{LOI_eva}
J.-L. Vay, D.~Sagan, A.~Huebl, M.~Th\'evenet, R.~Lehe, C.-K. Ng, H.~Vincenti,
  M.~Bussmann, A.~Debus, R.~Pausch, and J.~Qiang.
\newblock {End-to-End Virtual Accelerators (EVA)}.
\newblock \emph{Snowmass21 LOI}, 2020.
\newblock URL
  \url{https://www.snowmass21.org/docs/files/summaries/CompF/SNOWMASS21-CompF2_CompF0-AF1_AF0_Vay-067.pdf}.

\bibitem[Lehe et~al.(2020{\natexlab{a}})]{LOI_ML}
R.~Lehe et~al.
\newblock {Machine learning and surrogates models for simulation-based
  optimization of accelerator design}.
\newblock \emph{Snowmass21 LOI}, 2020{\natexlab{a}}.
\newblock URL
  \url{https://www.snowmass21.org/docs/files/summaries/CompF/SNOWMASS21-CompF2_CompF3-AF1_AF6_Lehe-075.pdf}.

\bibitem[Winklehner and Adelmann(2020)]{LOI_ML2}
D.~Winklehner and A.~Adelmann.
\newblock {Application of Machine Learning to Particle Accelerator
  Simulations}.
\newblock \emph{Snowmass21 LOI}, 2020.
\newblock URL
  \url{https://www.snowmass21.org/docs/files/summaries/CompF/SNOWMASS21-CompF3_CompF0-AF1_AF0_Winklehner-108.pdf}.

\bibitem[Nagaitsev et~al.(2020)Nagaitsev, Huang, J., Vay, Piot, Spentzouris,
  Rosenzweig, Cai, Cousineau, Conde, Hogan, Valishev, Minty, Zolkin, Huang,
  Shiltsev, Seeman, Byrd, and Patterson]{LOI_ABPRoadmap}
S.~Nagaitsev, Z.~Huang, Power J., J.-L. Vay, P.~Piot, L.~Spentzouris,
  J.~Rosenzweig, Y~Cai, S.~Cousineau, M.~Conde, M.~Hogan, A.~Valishev,
  M.~Minty, T.~Zolkin, X.~Huang, V.~Shiltsev, J.~Seeman, J.~Byrd, and J.R.
  Patterson.
\newblock {Accelerator and Beam Physics: Grand Challenges and Research
  Opportunities}.
\newblock \emph{Snowmass21 LOI}, 2020.
\newblock URL
  \url{https://www.snowmass21.org/docs/files/summaries/AF/SNOWMASS21-AF1_AF7_S_Nagaitsev-056.pdf}.

\bibitem[Adelmann(2019)]{adelmann:surrogate1}
Andreas. Adelmann.
\newblock On {Nonintrusive} {Uncertainty} {Quantification} and {Surrogate}
  {Model} {Construction} in {Particle} {Accelerator} {Modeling}.
\newblock \emph{SIAM/ASA Journal on Uncertainty Quantification}, 7\penalty0
  (2):\penalty0 383--416, January 2019.
\newblock \doi{10.1137/16M1061928}.
\newblock URL \url{https://epubs.siam.org/doi/10.1137/16M1061928}.

\bibitem[Van Der~Veken et~al.(2020)Van Der~Veken, Azzopardi, Blanc, Coyle, Fol,
  Giovannozzi, Pieloni, Redaelli, Salvachua~Ferrando, Schenk, Tomas~Garcia, and
  Valentino]{van_der_veken:ml1}
Frederik Van Der~Veken, Gabriella Azzopardi, Fred Blanc, Loic Coyle, Elena Fol,
  Massimo Giovannozzi, Tatiana Pieloni, Stefano Redaelli, Belen~Maria
  Salvachua~Ferrando, Michael Schenk, Rogelio Tomas~Garcia, and Gianluca
  Valentino.
\newblock Machine learning in accelerator physics: applications at the {CERN}
  {Large} {Hadron} {Collider}.
\newblock In \emph{Proceedings of {Artificial} {Intelligence} for {Science},
  {Industry} and {Society} {PoS}({AISIS2019})}, volume 372, page 044. SISSA
  Medialab, July 2020.
\newblock URL \url{https://pos.sissa.it/372/044/}.

\bibitem[Edelen et~al.(2016)Edelen, Biedron, Milton, and Edelen]{edelen:ml1}
A.~L. Edelen, S.~G. Biedron, S.~V. Milton, and J.~P. Edelen.
\newblock First {Steps} {Toward} {Incorporating} {Image} {Based} {Diagnostics}
  {Into} {Particle} {Accelerator} {Control} {Systems} {Using} {Convolutional}
  {Neural} {Networks}.
\newblock \emph{arXiv:1612.05662 [physics]}, December 2016.
\newblock URL \url{http://arxiv.org/abs/1612.05662}.
\newblock arXiv: 1612.05662.

\bibitem[Edelen et~al.(2020)Edelen, Neveu, Frey, Huber, Mayes, and
  Adelmann]{edelen:ml2}
Auralee Edelen, Nicole Neveu, Matthias Frey, Yannick Huber, Christopher Mayes,
  and Andreas Adelmann.
\newblock Machine learning for orders of magnitude speedup in multiobjective
  optimization of particle accelerator systems.
\newblock \emph{Physical Review Accelerators and Beams}, 23\penalty0
  (4):\penalty0 044601, April 2020.
\newblock \doi{10.1103/PhysRevAccelBeams.23.044601}.
\newblock URL
  \url{https://link.aps.org/doi/10.1103/PhysRevAccelBeams.23.044601}.
\newblock Publisher: American Physical Society.

\bibitem[Kirschner et~al.(2019{\natexlab{a}})Kirschner, Mutny, Hiller,
  Ischebeck, and Krause]{kirschner:bayesian1}
Johannes Kirschner, Mojmir Mutny, Nicole Hiller, Rasmus Ischebeck, and Andreas
  Krause.
\newblock Adaptive and {Safe} {Bayesian} {Optimization} in {High} {Dimensions}
  via {One}-{Dimensional} {Subspaces}.
\newblock \emph{arXiv:1902.03229 [cs, stat]}, May 2019{\natexlab{a}}.
\newblock URL \url{http://arxiv.org/abs/1902.03229}.
\newblock arXiv: 1902.03229.

\bibitem[Kirschner et~al.(2019{\natexlab{b}})Kirschner, Nonnenmacher, Mutny,
  Krause, Hiller, Ischebeck, and Adelmann]{kirschner:bayesian2}
Johannes Kirschner, Manuel Nonnenmacher, Mojmir Mutny, Andreas Krause, Nicole
  Hiller, Rasmus Ischebeck, and Andreas Adelmann.
\newblock Bayesian {Optimisation} for {Fast} and {Safe} {Parameter} {Tuning} of
  {SwissFEL}.
\newblock In \emph{{FEL2019}, {Proceedings} of the 39th {International}
  {Free}-{Electron} {Laser} {Conference}}, pages 707--710. JACoW Publishing,
  November 2019{\natexlab{b}}.
\newblock ISBN 978-3-95450-210-3.
\newblock \doi{10.3929/ethz-b-000385955}.
\newblock URL
  \url{https://www.research-collection.ethz.ch/handle/20.500.11850/385955}.
\newblock Accepted: 2019-12-17T07:59:07Z.

\bibitem[Duris et~al.(2020)Duris, Kennedy, Hanuka, Shtalenkova, Edelen,
  Baxevanis, Egger, Cope, McIntire, Ermon, and Ratner]{duris:bayesian1}
J.~Duris, D.~Kennedy, A.~Hanuka, J.~Shtalenkova, A.~Edelen, P.~Baxevanis,
  A.~Egger, T.~Cope, M.~McIntire, S.~Ermon, and D.~Ratner.
\newblock Bayesian {Optimization} of a {Free}-{Electron} {Laser}.
\newblock \emph{Physical Review Letters}, 124\penalty0 (12):\penalty0 124801,
  March 2020.
\newblock \doi{10.1103/PhysRevLett.124.124801}.
\newblock URL \url{https://link.aps.org/doi/10.1103/PhysRevLett.124.124801}.
\newblock Publisher: American Physical Society.

\bibitem[Bussmann et~al.(2013)Bussmann, Burau, Cowan, Debus, Huebl, Juckeland,
  Kluge, Nagel, Pausch, Schmitt, Schramm, Schuchart, and Widera]{PIConGPU2013}
M.~Bussmann, H.~Burau, T.~E. Cowan, A.~Debus, A.~Huebl, G.~Juckeland, T.~Kluge,
  W.~E. Nagel, R.~Pausch, F.~Schmitt, U.~Schramm, J.~Schuchart, and R.~Widera.
\newblock Radiative signatures of the relativistic kelvin-helmholtz
  instability.
\newblock In \emph{Proceedings of the International Conference on High
  Performance Computing, Networking, Storage and Analysis}, SC '13, pages
  5:1--5:12, New York, NY, USA, 2013. ACM.
\newblock ISBN 978-1-4503-2378-9.
\newblock \doi{10.1145/2503210.2504564}.
\newblock URL \url{http://doi.acm.org/10.1145/2503210.2504564}.

\bibitem[Myers et~al.(2021)Myers, Almgren, Amorim, Bell, Fedeli, Ge, Gott,
  Grote, Hogan, Huebl, Jambunathan, Lehe, Ng, Rowan, Shapoval, Thévenet, Vay,
  Vincenti, Yang, Zaim, Zhang, Zhao, and Zoni]{Myers2021}
A.~Myers, A.~Almgren, L.~D. Amorim, J.~Bell, L.~Fedeli, L.~Ge, K.~Gott, D.~P.
  Grote, M.~Hogan, A.~Huebl, R.~Jambunathan, R.~Lehe, C.~Ng, M.~Rowan,
  O.~Shapoval, M.~Thévenet, J.~L. Vay, H.~Vincenti, E.~Yang, N.~Zaim,
  W.~Zhang, Y.~Zhao, and E.~Zoni.
\newblock Porting warpx to gpu-accelerated platforms, 2021.

\bibitem[kok()]{kokkos}
{Kokkos {\textperiodcentered} GitHub}.
\newblock URL \url{https://github.com/kokkos}.

\bibitem[Zenker et~al.(2016)Zenker, Worpitz, Widera, Huebl, Juckeland,
  Kn{\"{u}}pfer, Nagel, and Bussmann]{Alpaka}
Erik Zenker, Benjamin Worpitz, René Widera, Axel Huebl, Guido Juckeland,
  Andreas Kn{\"{u}}pfer, Wolfgang~E. Nagel, and Michael Bussmann.
\newblock {Alpaka - An Abstraction Library for Parallel Kernel Acceleration}.
\newblock \emph{Proceedings - 2016 IEEE 30th International Parallel and
  Distributed Processing Symposium, IPDPS 2016}, pages 631--640, 2 2016.
\newblock \doi{10.1109/ipdpsw.2016.50}.
\newblock URL \url{https://arxiv.org/abs/1602.08477v1}.

\bibitem[Raj()]{Raja}
{GitHub - LLNL/RAJA: RAJA Performance Portability Layer (C++)}.
\newblock URL \url{https://github.com/LLNL/RAJA}.

\bibitem[AMR()]{AMReX}
{AMReX}.
\newblock URL \url{https://amrex-codes.github.io/}.

\bibitem[{The Copa Team}()]{Copa}
{The Copa Team}.
\newblock {Copa}.
\newblock URL \url{https://github.com/ECP-copa}.

\bibitem[Godfrey(1974)]{GodfreyJCP1974}
Brendan~B Godfrey.
\newblock \emph{Journal of Computational Physics}, 15\penalty0 (4):\penalty0
  504 -- 521, 1974.
\newblock ISSN 0021-9991.
\newblock \doi{10.1016/0021-9991(74)90076-X}.
\newblock URL
  \url{http://www.sciencedirect.com/science/article/pii/002199917490076X}.

\bibitem[Godfrey(1975)]{GodfreyJCP1975}
Brendan~B. Godfrey.
\newblock Canonical momenta and numerical instabilities in particle codes.
\newblock \emph{Journal of Computational Physics}, 19\penalty0 (1):\penalty0 58
  -- 76, 1975.
\newblock ISSN 0021-9991.
\newblock \doi{10.1016/0021-9991(75)90116-3}.
\newblock URL
  \url{http://www.sciencedirect.com/science/article/pii/0021999175901163}.

\bibitem[Vay {\textbackslash}it Et~Al.(2009)]{Vaypac09}
J.-L. Vay {\textbackslash}it Et~Al.
\newblock {Application Of The Reduction Of Scale Range In A Lorentz Boosted
  Frame To The Numerical Simulation Of Particle Acceleration Devices}.
\newblock In \emph{Proc. Particle Accelerator Conference}, Vancouver, Canada,
  2009.

\bibitem[Martins et~al.(2010)Martins, Fonseca, Silva, Lu, and
  Mori]{Martinscpc10}
Samuel~F Martins, Ricardo~A Fonseca, Luis~O Silva, Wei Lu, and Warren~B Mori.
\newblock {Numerical Simulations Of Laser Wakefield Accelerators In Optimal
  Lorentz Frames}.
\newblock \emph{Computer Physics Communications}, 181\penalty0 (5):\penalty0
  869--875, 5 2010.
\newblock ISSN 0010-4655.
\newblock \doi{10.1016/J.Cpc.2009.12.023}.

\bibitem[Vay et~al.(2011{\natexlab{a}})Vay, Geddes, Cormier-Michel, and
  Grote]{VayJCP2011}
J~L Vay, C~G~R Geddes, E~Cormier-Michel, and D~P Grote.
\newblock {Numerical Methods For Instability Mitigation In The Modeling Of
  Laser Wakefield Accelerators In A Lorentz-Boosted Frame}.
\newblock \emph{Journal of Computational Physics}, 230\penalty0 (15):\penalty0
  5908--5929, 7 2011{\natexlab{a}}.
\newblock \doi{10.1016/J.Jcp.2011.04.003}.

\bibitem[Vay et~al.(2011{\natexlab{b}})Vay, Geddes, Cormier-Michel, and
  Grote]{VayPOPL2011}
Jl~Vay, C~G~R Geddes, E~Cormier-Michel, and D~P Grote.
\newblock {Effects Of Hyperbolic Rotation In Minkowski Space On The Modeling Of
  Plasma Accelerators In A Lorentz Boosted Frame}.
\newblock \emph{Physics Of Plasmas}, 18\penalty0 (3):\penalty0 30701, 3
  2011{\natexlab{b}}.
\newblock \doi{10.1063/1.3559483}.

\bibitem[Godfrey and Vay(2013)]{GodfreyJCP2013}
Brendan~B Godfrey and Jean-Luc Vay.
\newblock {Numerical stability of relativistic beam multidimensional
  {\{}PIC{\}} simulations employing the Esirkepov algorithm}.
\newblock \emph{Journal of Computational Physics}, 248\penalty0 (0):\penalty0
  33--46, 2013.
\newblock ISSN 0021-9991.
\newblock \doi{http://dx.doi.org/10.1016/j.jcp.2013.04.006}.
\newblock URL
  \url{http://www.sciencedirect.com/science/article/pii/S0021999113002556}.

\bibitem[Xu et~al.(2013)Xu, Yu, Martins, Tsung, Decyk, Vieira, Fonseca, Lu,
  Silva, and Mori]{XuCPC2013}
Xinlu Xu, Peicheng Yu, Samual~F Martins, Frank~S Tsung, Viktor~K Decyk, Jorge
  Vieira, Ricardo~A Fonseca, Wei Lu, Luis~O Silva, and Warren~B Mori.
\newblock {Numerical instability due to relativistic plasma drift in EM-PIC
  simulations}.
\newblock \emph{Computer Physics Communications}, 184\penalty0 (11):\penalty0
  2503--2514, 2013.
\newblock ISSN 0010-4655.
\newblock \doi{http://dx.doi.org/10.1016/j.cpc.2013.07.003}.
\newblock URL
  \url{http://www.sciencedirect.com/science/article/pii/S0010465513002312}.

\bibitem[Godfrey et~al.(2014{\natexlab{a}})Godfrey, Vay, and
  Haber]{GodfreyJCP2014}
Brendan~B Godfrey, Jean-Luc Vay, and Irving Haber.
\newblock {Numerical stability analysis of the pseudo-spectral analytical
  time-domain {\{}PIC{\}} algorithm}.
\newblock \emph{Journal of Computational Physics}, 258\penalty0 (0):\penalty0
  689--704, 2014{\natexlab{a}}.
\newblock ISSN 0021-9991.
\newblock \doi{http://dx.doi.org/10.1016/j.jcp.2013.10.053}.
\newblock URL
  \url{http://www.sciencedirect.com/science/article/pii/S0021999113007298}.

\bibitem[Godfrey and Vay(2014)]{GodfreyJCP2014b}
Brendan~B. Godfrey and Jean-Luc Vay.
\newblock Suppressing the numerical cherenkov instability in \{FDTD\} \{PIC\}
  codes.
\newblock \emph{Journal of Computational Physics}, 267:\penalty0 1 -- 6, 2014.
\newblock ISSN 0021-9991.
\newblock \doi{10.1016/j.jcp.2014.02.022}.
\newblock URL
  \url{http://www.sciencedirect.com/science/article/pii/S0021999114001429}.

\bibitem[Godfrey et~al.(2014{\natexlab{b}})Godfrey, Vay, and
  Haber]{GodfreyIEEE2014}
Brendan~B. Godfrey, Jean~Luc Vay, and Irving Haber.
\newblock {Numerical stability improvements for the pseudospectral EM PIC
  algorithm}.
\newblock \emph{IEEE Transactions on Plasma Science}, 42\penalty0 (5):\penalty0
  1339--1344, 2014{\natexlab{b}}.

\bibitem[Godfrey and Vay(2015)]{GodfreyCPC2015}
Brendan~B. Godfrey and Jean~Luc Vay.
\newblock {Improved numerical Cherenkov instability suppression in the
  generalized PSTD PIC algorithm}.
\newblock \emph{Computer Physics Communications}, 196:\penalty0 221--225, 2015.

\bibitem[Yu et~al.(2015{\natexlab{a}})Yu, Xu, Decyk, Fiuza, Vieira, Tsung,
  Fonseca, Lu, Silva, and Mori]{YuCPC2015}
Peicheng Yu, Xinlu Xu, Viktor~K. Decyk, Frederico Fiuza, Jorge Vieira, Frank~S.
  Tsung, Ricardo~A. Fonseca, Wei Lu, Luis~O. Silva, and Warren~B. Mori.
\newblock {Elimination of the numerical Cerenkov instability for spectral
  EM-PIC codes}.
\newblock \emph{Computer Physics Communications}, 192:\penalty0 32--47, 7
  2015{\natexlab{a}}.
\newblock ISSN 00104655.
\newblock \doi{10.1016/j.cpc.2015.02.018}.
\newblock URL
  \url{https://apps.webofknowledge.com/full_record.do?product=UA&search_mode=GeneralSearch&qid=2&SID=1CanLFIHrQ5v8O7cxqV&page=1&doc=3}.

\bibitem[Yu et~al.(2015{\natexlab{b}})Yu, Xu, Tableman, Decyk, Tsung, Fiuza,
  Davidson, Vieira, Fonseca, Lu, Silva, and Mori]{YuCPC2015-Circ}
Peicheng Yu, Xinlu Xu, Adam Tableman, Viktor~K. Decyk, Frank~S. Tsung,
  Frederico Fiuza, Asher Davidson, Jorge Vieira, Ricardo~A. Fonseca, Wei Lu,
  Luis~O. Silva, and Warren~B. Mori.
\newblock {Mitigation of numerical Cerenkov radiation and instability using a
  hybrid finite difference-FFT Maxwell solver and a local charge conserving
  current deposit}.
\newblock \emph{Computer Physics Communications}, 197:\penalty0 144--152, 12
  2015{\natexlab{b}}.
\newblock ISSN 00104655.
\newblock \doi{10.1016/j.cpc.2015.08.026}.
\newblock URL
  \url{https://apps.webofknowledge.com/full_record.do?product=UA&search_mode=GeneralSearch&qid=2&SID=1CanLFIHrQ5v8O7cxqV&page=1&doc=2}.

\bibitem[Lehe et~al.(2016{\natexlab{a}})Lehe, Kirchen, Godfrey, Maier, and
  Vay]{LehePRE2016}
Remi Lehe, Manuel Kirchen, Brendan~B. Godfrey, Andreas~R. Maier, and Jean-Luc
  Vay.
\newblock {Elimination of numerical Cherenkov instability in flowing-plasma
  particle-in-cell simulations by using Galilean coordinates}.
\newblock \emph{Physical Review E}, 94\penalty0 (5):\penalty0 053305, 11
  2016{\natexlab{a}}.
\newblock ISSN 2470-0045.
\newblock \doi{10.1103/PhysRevE.94.053305}.
\newblock URL \url{https://link.aps.org/doi/10.1103/PhysRevE.94.053305}.

\bibitem[Kirchen et~al.(2016)Kirchen, Lehe, Godfrey, Dornmair, Jalas, Peters,
  Vay, and Maier]{KirchenPoP2016}
M.~Kirchen, R.~Lehe, B.~B. Godfrey, I.~Dornmair, S.~Jalas, K.~Peters, J.-L.
  Vay, and A.~R. Maier.
\newblock {Stable discrete representation of relativistically drifting
  plasmas}.
\newblock \emph{Physics of Plasmas}, 23\penalty0 (10):\penalty0 100704, 10
  2016.
\newblock ISSN 1070-664X.
\newblock \doi{10.1063/1.4964770}.
\newblock URL \url{http://aip.scitation.org/doi/10.1063/1.4964770}.

\bibitem[Kirchen et~al.(2020)Kirchen, Lehe, Jalas, Shapoval, Vay, and
  Maier]{KirchenPRE2020}
Manuel Kirchen, Remi Lehe, Soeren Jalas, Olga Shapoval, Jean~Luc Vay, and
  Andreas~R. Maier.
\newblock {Scalable spectral solver in Galilean coordinates for eliminating the
  numerical Cherenkov instability in particle-in-cell simulations of streaming
  plasmas}.
\newblock \emph{Physical Review E}, 102\penalty0 (1):\penalty0 13202, 7 2020.
\newblock ISSN 24700053.
\newblock \doi{10.1103/PhysRevE.102.013202}.
\newblock URL
  \url{https://journals.aps.org/pre/abstract/10.1103/PhysRevE.102.013202}.

\bibitem[Li et~al.(2017)Li, Yu, Xu, Fiuza, Decyk, Dalichaouch, Davidson,
  Tableman, An, Tsung, Fonseca, Lu, and Mori]{FeiCPC2017}
Fei Li, Peicheng Yu, Xinlu Xu, Frederico Fiuza, Viktor~K. Decyk, Thamine
  Dalichaouch, Asher Davidson, Adam Tableman, Weiming An, Frank~S. Tsung,
  Ricardo~A. Fonseca, Wei Lu, and Warren~B. Mori.
\newblock Controlling the numerical cerenkov instability in pic simulations
  using a customized finite difference maxwell solver and a local fft based
  current correction.
\newblock \emph{Computer Physics Communications}, 214:\penalty0 6--17, 2017.
\newblock ISSN 0010-4655.
\newblock \doi{https://doi.org/10.1016/j.cpc.2017.01.001}.
\newblock URL
  \url{https://www.sciencedirect.com/science/article/pii/S0010465517300012}.

\bibitem[Pukhov(2020)]{PukhovJCP2020}
Alexander Pukhov.
\newblock {X-dispersionless Maxwell solver for plasma-based particle
  acceleration}.
\newblock \emph{Journal of Computational Physics}, 418:\penalty0 109622, 10
  2020.
\newblock ISSN 0021-9991.
\newblock \doi{10.1016/J.JCP.2020.109622}.

\bibitem[Haber et~al.(1973)Haber, Lee, Klein, and Boris]{Habericnsp73}
I~Haber, R~Lee, Hh~Klein, and Jp~Boris.
\newblock {Advances In Electromagnetic Simulation Techniques}.
\newblock In \emph{Proc. Sixth Conf. Num. Sim. Plasmas}, pages 46--48,
  Berkeley, Ca, 1973.

\bibitem[Vay et~al.(2013)Vay, Haber, and Godfrey]{VayJCP2013}
Jean~Luc Vay, Irving Haber, and Brendan~B. Godfrey.
\newblock {A domain decomposition method for pseudo-spectral electromagnetic
  simulations of plasmas}.
\newblock \emph{Journal of Computational Physics}, 243:\penalty0 260--268,
  2013.

\bibitem[Vincenti and Vay(2016)]{VincentiCPC2016}
H.~Vincenti and J.-L. Vay.
\newblock {Detailed analysis of the effects of stencil spatial variations with
  arbitrary high-order finite-difference Maxwell solver}.
\newblock \emph{Computer Physics Communications}, 200:\penalty0 147--167, mar
  2016.
\newblock ISSN 00104655.
\newblock \doi{10.1016/j.cpc.2015.11.009}.
\newblock URL
  \url{https://apps.webofknowledge.com/full{\_}record.do?product=UA{\&}search{\_}mode=GeneralSearch{\&}qid=1{\&}SID=1CanLFIHrQ5v8O7cxqV{\&}page=1{\&}doc=2}.

\bibitem[Jalas et~al.(2017)Jalas, Dornmair, Lehe, Vincenti, Vay, Kirchen, and
  Maier]{JalasPoP2017}
S.~Jalas, I.~Dornmair, R.~Lehe, H.~Vincenti, J.-L. Vay, M.~Kirchen, and A.~R.
  Maier.
\newblock {Accurate modeling of plasma acceleration with arbitrary order
  pseudo-spectral particle-in-cell methods}.
\newblock \emph{Physics of Plasmas}, 24\penalty0 (3):\penalty0 033115, mar
  2017.
\newblock ISSN 1070-664X.
\newblock \doi{10.1063/1.4978569}.
\newblock URL \url{http://aip.scitation.org/doi/10.1063/1.4978569}.

\bibitem[Shapoval et~al.(2021)Shapoval, Lehe, Th{\'e}venet, Zoni, Zhao, and
  Vay]{ShapovalARXIV2021}
Olga Shapoval, Remi Lehe, Maxence Th{\'e}venet, Edoardo Zoni, Yinjian Zhao, and
  Jean-Luc Vay.
\newblock {Overcoming timestep limitations in boosted-frame Particle-In-Cell
  simulations of plasma-based acceleration}.
\newblock \emph{arXiv preprint arXiv:2104.13995}, 2021.

\bibitem[Zon()]{ZoniARXIV2021}
{A Hybrid Nodal-Staggered Pseudo-Spectral Electromagnetic Particle-In-Cell
  Method with Finite-Order Centering}, author = {Zoni, Edoardo and Lehe, Remi
  and Shapoval, Olga and Belkin, Daniel and Zaïm, Neil and Fedeli, Luca and
  Vincenti, Henri and Vay, Jean-Luc}, journal = {arXiv preprint arXiv:0.0},
  year = {2021}.

\bibitem[Vay et~al.(2004{\natexlab{a}})Vay, Colella, Kwan, McCorquodale,
  Serafini, Friedman, Grote, Westenskow, Adam, H{\'{e}}ron, and Haber]{Vay2004}
J.-L. Vay, P.~Colella, J.W. Kwan, P.~McCorquodale, D.B. Serafini, A.~Friedman,
  D.P. Grote, G.~Westenskow, J.-C. Adam, A.~H{\'{e}}ron, and I.~Haber.
\newblock {Application of adaptive mesh refinement to particle-in-cell
  simulations of plasmas and beams}.
\newblock \emph{Physics of Plasmas}, 11\penalty0 (5 PART 2),
  2004{\natexlab{a}}.
\newblock ISSN 1070664X.
\newblock \doi{10.1063/1.1689669}.

\bibitem[Vay et~al.(2004{\natexlab{b}})Vay, Adam, and H{\'{e}}ron]{VayCPC2004}
J.-L. Vay, J.-C. Adam, and A.~H{\'{e}}ron.
\newblock {Asymmetric PML for the absorption of waves. Application to mesh
  refinement in electromagnetic Particle-In-Cell plasma simulations}.
\newblock \emph{Computer Physics Communications}, 164\penalty0 (1-3),
  2004{\natexlab{b}}.
\newblock ISSN 00104655.
\newblock \doi{10.1016/j.cpc.2004.06.026}.

\bibitem[Vay et~al.(2012{\natexlab{b}})Vay, Grote, Cohen, and
  Friedman]{VayCSD2012}
J.-L. Vay, D.P. Grote, R.H. Cohen, and A.~Friedman.
\newblock {Novel methods in the Particle-In-Cell accelerator Code-Framework
  Warp}.
\newblock \emph{Computational Science and Discovery}, 5\penalty0 (1),
  2012{\natexlab{b}}.
\newblock ISSN 17494680.
\newblock \doi{10.1088/1749-4699/5/1/014019}.

\bibitem[Vay et~al.(2018)Vay, Almgren, Bell, Ge, Grote, Hogan, Kononenko, Lehe,
  Myers, Ng, Park, Ryne, Shapoval, Th{\'{e}}venet, and Zhang]{WarpXEAAC2017}
J.-L. Vay, A.~Almgren, J.~Bell, L.~Ge, D.P. Grote, M.~Hogan, O.~Kononenko,
  R.~Lehe, A.~Myers, C.~Ng, J.~Park, R.~Ryne, O.~Shapoval, M.~Th{\'{e}}venet,
  and W.~Zhang.
\newblock {Warp-X: A new exascale computing platform for beam–plasma
  simulations}.
\newblock \emph{Nuclear Instruments and Methods in Physics Research Section A:
  Accelerators, Spectrometers, Detectors and Associated Equipment}, 1 2018.
\newblock ISSN 0168-9002.
\newblock \doi{10.1016/J.NIMA.2018.01.035}.
\newblock URL
  \url{https://www.sciencedirect.com/science/article/pii/S0168900218300524}.

\bibitem[Vay et~al.(2019)Vay, Almgren, Bell, Lehe, Myers, Park, Shapoval,
  Thevenet, Zhang, Grote, Hogan, Ge, and Ng]{WarpXAAC2018}
J.~L. Vay, A.~Almgren, J.~Bell, R.~Lehe, A.~Myers, J.~Park, O.~Shapoval,
  M.~Thevenet, W.~Zhang, D.~P. Grote, M.~Hogan, L.~Ge, and C.~Ng.
\newblock {Toward plasma wakefield simulations at exascale}.
\newblock In \emph{2018 IEEE Advanced Accelerator Concepts Workshop, ACC 2018 -
  Proceedings}. Institute of Electrical and Electronics Engineers Inc., 3 2019.
\newblock ISBN 9781538677216.
\newblock \doi{10.1109/AAC.2018.8659392}.

\bibitem[Lehe et~al.(2016{\natexlab{b}})Lehe, Kirchen, Andriyash, Godfrey, and
  Vay]{Lehe2016}
Rémi Lehe, Manuel Kirchen, Igor~A. Andriyash, Brendan~B. Godfrey, and Jean-Luc
  Vay.
\newblock {A spectral, quasi-cylindrical and dispersion-free Particle-In-Cell
  algorithm}.
\newblock \emph{Computer Physics Communications}, 203:\penalty0 66--82,
  2016{\natexlab{b}}.
\newblock ISSN 00104655.
\newblock \doi{10.1016/j.cpc.2016.02.007}.

\bibitem[Li et~al.(2021)Li, An, Decyk, Xu, Hogan, and Mori]{LiCPC2021}
Fei Li, Weiming An, Viktor~K. Decyk, Xinlu Xu, Mark~J. Hogan, and Warren~B.
  Mori.
\newblock {A quasi-static particle-in-cell algorithm based on an azimuthal
  Fourier decomposition for highly efficient simulations of plasma-based
  acceleration: QPAD}.
\newblock \emph{Computer Physics Communications}, 261:\penalty0 107784, 4 2021.
\newblock ISSN 0010-4655.
\newblock \doi{10.1016/J.CPC.2020.107784}.

\bibitem[FBP()]{FBPIC}
{FBPIC}.
\newblock URL \url{https://fbpic.github.io}.

\bibitem[Kallala et~al.(2019)Kallala, Vay, and Vincenti]{Kallala2019}
Haithem Kallala, Jean~Luc Vay, and Henri Vincenti.
\newblock {A generalized massively parallel ultra-high order FFT-based Maxwell
  solver}.
\newblock \emph{Computer Physics Communications}, 244:\penalty0 25--34, 11
  2019.
\newblock ISSN 00104655.
\newblock \doi{10.1016/j.cpc.2019.07.009}.

\bibitem[Blaclard et~al.(2017)Blaclard, Vincenti, Lehe, and Vay]{Blaclard2017}
G.~Blaclard, H.~Vincenti, R.~Lehe, and J.~L. Vay.
\newblock {Pseudospectral Maxwell solvers for an accurate modeling of Doppler
  harmonic generation on plasma mirrors with particle-in-cell codes}.
\newblock \emph{Physical Review E}, 96\penalty0 (3):\penalty0 033305, 9 2017.
\newblock ISSN 2470-0045.
\newblock \doi{10.1103/PhysRevE.96.033305}.
\newblock URL \url{https://link.aps.org/doi/10.1103/PhysRevE.96.033305}.

\bibitem[Leblanc et~al.(2017)Leblanc, Monchoc{\'{e}}, Vincenti, Kahaly, Vay,
  and Qu{\'{e}}r{\'{e}}]{Leblanc2017}
A.~Leblanc, S.~Monchoc{\'{e}}, H.~Vincenti, S.~Kahaly, J.-L. Vay, and
  F.~Qu{\'{e}}r{\'{e}}.
\newblock {Spatial Properties of High-Order Harmonic Beams from Plasma Mirrors:
  A Ptychographic Study}.
\newblock \emph{Physical Review Letters}, 119\penalty0 (15):\penalty0 155001,
  10 2017.
\newblock ISSN 0031-9007.
\newblock \doi{10.1103/PhysRevLett.119.155001}.
\newblock URL \url{https://link.aps.org/doi/10.1103/PhysRevLett.119.155001}.

\bibitem[Vincenti(2019)]{Vincenti2019}
Henri Vincenti.
\newblock {Achieving Extreme Light Intensities using Optically Curved
  Relativistic Plasma Mirrors}.
\newblock \emph{Physical Review Letters}, 123\penalty0 (10):\penalty0 105001, 9
  2019.
\newblock ISSN 10797114.
\newblock \doi{10.1103/PhysRevLett.123.105001}.

\bibitem[Chopineau et~al.(2019)Chopineau, Leblanc, Blaclard, Denoeud,
  Th{\'{e}}venet, Vay, Bonnaud, Martin, Vincenti, and
  Qu{\'{e}}r{\'{e}}]{Chopineau2019}
L.~Chopineau, A.~Leblanc, G.~Blaclard, A.~Denoeud, M.~Th{\'{e}}venet, J.~L.
  Vay, G.~Bonnaud, Ph~Martin, H.~Vincenti, and F.~Qu{\'{e}}r{\'{e}}.
\newblock {Identification of Coupling Mechanisms between Ultraintense Laser
  Light and Dense Plasmas}.
\newblock \emph{Physical Review X}, 9\penalty0 (1):\penalty0 011050, 3 2019.
\newblock ISSN 21603308.
\newblock \doi{10.1103/PhysRevX.9.011050}.

\bibitem[Fedeli et~al.(2021)Fedeli, Sainte-Marie, Zaim, Th{\'{e}}venet, Vay,
  Myers, Qu{\'{e}}r{\'{e}}, and Vincenti]{FedeliPRL2021}
L.~Fedeli, A.~Sainte-Marie, N.~Zaim, M.~Th{\'{e}}venet, J.~L. Vay, A.~Myers,
  F.~Qu{\'{e}}r{\'{e}}, and H.~Vincenti.
\newblock {Probing strong-field QED with Doppler-boosted petawatt-class lasers
  (accepted paper)}.
\newblock \emph{Physical Review Letters}, 2021.
\newblock URL
  \url{https://journals.aps.org/prl/accepted/7f072Y26M8d1616c368b4ab2c339eb42b2eb19f58}.

\bibitem[Wan et~al.(2021)Wan, Huebl, Gu, Poeschel, Gainaru, Wang, Chen, Liang,
  Ganyushin, Munson, Foster, Vay, Podhorszki, Wu, and Klasky]{Wan2021}
Lipeng Wan, Axel Huebl, Junmin Gu, Franz Poeschel, Ana Gainaru, Ruonan Wang,
  Jieyang Chen, Xin Liang, Dmitry Ganyushin, Todd Munson, Ian Foster, Jean-Luc
  Vay, Norbert Podhorszki, Kesheng Wu, and Scott Klasky.
\newblock {Improving I/O Performance for Exascale Applications through Online
  Data Layout Reorganization}.
\newblock \emph{accepted in IEEE Transactions on Parallel and Distributed
  Systems}, 2021.
\newblock URL \url{https://arxiv.org/abs/2107.07108}.

\bibitem[Huebl et~al.(2017)Huebl, Widera, Schmitt, Matthes, Podhorszki, Choi,
  Klasky, and Bussmann]{Huebl2017}
Axel Huebl, Ren{\'e} Widera, Felix Schmitt, Alexander Matthes, Norbert
  Podhorszki, Jong~Youl Choi, Scott Klasky, and Michael Bussmann.
\newblock On the scalability of data reduction techniques in current and
  upcoming hpc systems from an application perspective.
\newblock In Julian~M. Kunkel, Rio Yokota, Michela Taufer, and John Shalf,
  editors, \emph{High Performance Computing}, pages 15--29, Cham, 2017.
  Springer International Publishing.
\newblock ISBN 978-3-319-67630-2.

\bibitem[Huebl et~al.(2015)Huebl, Lehe, Vay, Grote, Sbalzarini, Kuschel, Sagan,
  Pérez, Koller, and Bussmann]{openPMD}
Axel Huebl, Rémi Lehe, Jean-Luc Vay, David~P. Grote, Ivo Sbalzarini, Stephan
  Kuschel, David Sagan, Frédéric Pérez, Fabian Koller, and Michael Bussmann.
\newblock {openPMD: A meta data standard for particle and mesh based data}.
\newblock 2015.
\newblock \doi{10.5281/zenodo.591699}.
\newblock URL \url{https://www.openpmd.org}.

\bibitem[Poeschel et~al.(2021)Poeschel, E, Godoy, Podhorszki, Klasky,
  Eisenhauer, Davis, Wan, Gainaru, Gu, Koller, Widera, Bussmann, and
  Huebl]{Poeschel2021}
Franz Poeschel, Juncheng E, William~F. Godoy, Norbert Podhorszki, Scott Klasky,
  Greg Eisenhauer, Philip~E. Davis, Lipeng Wan, Ana Gainaru, Junmin Gu, Fabian
  Koller, Rene Widera, Michael Bussmann, and Axel Huebl.
\newblock {Transitioning from file-based HPC workflows to streaming data
  pipelines with openPMD and ADIOS2}.
\newblock \emph{submitted}, 2021.

\bibitem[CAM()]{CAMPA}
{CAMPA: Consortium for Advanced Modeling of Particle Accelerators}.
\newblock URL \url{http://campa.lbl.gov}.

\bibitem[PIC()]{PICMI}
{PICMI}.
\newblock URL \url{https://github.com/picmi-standard}.

\bibitem[Godoy et~al.(2020)Godoy, Podhorszki, Wang, Atkins, Eisenhauer, Gu,
  Davis, Choi, Germaschewski, Huck, Huebl, Kim, Kress, Kurc, Liu, Logan, Mehta,
  Ostrouchov, Parashar, Poeschel, Pugmire, Suchyta, Takahashi, Thompson,
  Tsutsumi, Wan, Wolf, Wu, and Klasky]{ADIOS2}
William~F. Godoy, Norbert Podhorszki, Ruonan Wang, Chuck Atkins, Greg
  Eisenhauer, Junmin Gu, Philip Davis, Jong Choi, Kai Germaschewski, Kevin
  Huck, Axel Huebl, Mark Kim, James Kress, Tahsin Kurc, Qing Liu, Jeremy Logan,
  Kshitij Mehta, George Ostrouchov, Manish Parashar, Franz Poeschel, David
  Pugmire, Eric Suchyta, Keichi Takahashi, Nick Thompson, Seiji Tsutsumi,
  Lipeng Wan, Matthew Wolf, Kesheng Wu, and Scott Klasky.
\newblock Adios 2: The adaptable input output system. a framework for
  high-performance data management.
\newblock \emph{SoftwareX}, 12:\penalty0 100561, 2020.
\newblock ISSN 2352-7110.
\newblock \doi{https://doi.org/10.1016/j.softx.2020.100561}.
\newblock URL
  \url{https://www.sciencedirect.com/science/article/pii/S2352711019302560}.

\bibitem[{The HDF Group}()]{HDF5}
{The HDF Group}.
\newblock {Hierarchical data format version 5}.
\newblock URL \url{http://www.hdfgroup.org/HDF5}.

\bibitem[Huebl et~al.(2018)Huebl, Poeschel, Koller, and Gu]{openPMDapi}
Axel Huebl, Franz Poeschel, Fabian Koller, and Junmin Gu.
\newblock {openPMD-api: C++ \& Python API for Scientific I/O with openPMD}.
\newblock 2018.
\newblock \doi{10.14278/rodare.27}.
\newblock URL \url{https://github.com/openPMD/openPMD-api}.

\bibitem[Fonseca {\textbackslash}It Et~Al.(2002)]{Osiris}
R~A Fonseca {\textbackslash}It Et~Al.
\newblock {No Title}.
\newblock \emph{Lec. Notes In Comp. Sci.}, 2329:\penalty0 342, 2002.

\bibitem[Vay et~al.(2021)Vay, Huebl, Almgren, Amorim, Bell, Fedeli, Ge, Gott,
  Grote, Hogan, Jambunathan, Lehe, Myers, Ng, Rowan, Shapoval, Th{\'{e}}venet,
  Vincenti, Yang, Za{\"{i}}m, Zhang, Zhao, and Zoni]{VayPoP2021}
J.~L. Vay, A.~Huebl, A.~Almgren, L.~D. Amorim, J.~Bell, L.~Fedeli, L.~Ge,
  K.~Gott, D.~P. Grote, M.~Hogan, R.~Jambunathan, R.~Lehe, A.~Myers, C.~Ng,
  M.~Rowan, O.~Shapoval, M.~Th{\'{e}}venet, H.~Vincenti, E.~Yang,
  N.~Za{\"{i}}m, W.~Zhang, Y.~Zhao, and E.~Zoni.
\newblock {Modeling of a chain of three plasma accelerator stages with the
  WarpX electromagnetic PIC code on GPUs}.
\newblock \emph{Physics of Plasmas}, 28\penalty0 (2):\penalty0 23105, 2 2021.
\newblock ISSN 10897674.
\newblock \doi{10.1063/5.0028512}.
\newblock URL \url{https://doi.org/10.1063/5.0028512}.

\bibitem[Th{\'e}venet et~al.(2021)Th{\'e}venet, Diederichs, Sinn, Lehe, Huebl,
  Myers, Zhang, and Vay]{HipacePP}
Maxence Th{\'e}venet, Severin Diederichs, Alexander Sinn, Remi Lehe, Axel
  Huebl, Andrew Myers, Weiqun Zhang, and Jean-Luc Vay.
\newblock {HiPACE++: Highly efficient Plasma Accelerator Emulation, quasistatic
  particle-in-cell code}, 2021.
\newblock URL \url{https://github.com/Hi-PACE/hipace}.

\bibitem[IDE()]{IDEAS}
{IDEAS: Interoperable Design of Extreme-scale Application Software}.
\newblock URL \url{https://ideas-productivity.org}.

\bibitem[xSD()]{xSDK}
{xSDK: Extreme-scale Scientific Software Development Kit}.
\newblock URL \url{http://xsdk.info}.

\bibitem[Lehe et~al.(2020{\natexlab{b}})Lehe, Huebl, Vay, Friedman, Thevenet,
  Mitchell, Bruhwiler, Grote, Cowan, Vincenti, Hanuka, Cros, Yoffe, Widera,
  Bussmann, and Edelen]{LOI_industry}
Remi Lehe, Axel Huebl, Jean-Luc Vay, Alexander Friedman, Maxence Thevenet, Chad
  Mitchell, David Bruhwiler, David Grote, Benjamin Cowan, Henri Vincenti, Adi
  Hanuka, Brigitte Cros, Samuel Yoffe, Rene Widera, Michael Bussmann, and
  Auralee~Linscott Edelen.
\newblock {Embracing modern software tools and user-friendly practices, when
  distributing scientific codes}.
\newblock \emph{Snowmass21 LOI}, 2020{\natexlab{b}}.
\newblock URL
  \url{https://www.snowmass21.org/docs/files/summaries/CompF/SNOWMASS21-CompF2_CompF0_Lehe-076.pdf}.

\bibitem[spa()]{spack}
Spack.
\newblock \url{https://spack.io/}.

\bibitem[con()]{conda}
Conda.
\newblock \url{https://docs.conda.io/en/latest/}.

\bibitem[pip()]{pip}
The phython package installer.
\newblock \url{https://pip.pypa.io/en/stable/}.

\end{thebibliography}

\end{document}